\documentclass[aps,prl,reprint,superscriptaddress]{revtex4-1}
\usepackage{amsmath,amssymb,ulem}
\usepackage{bbold}
\usepackage{float}
\usepackage{graphicx}
\usepackage{dcolumn}
\usepackage{bm}
\usepackage[colorlinks]{hyperref}

\begin{document}

\title{Broadband continuous variable entanglement generation using\\ Kerr-free Josephson metamaterial}

\author{M. R. Perelshtein}
    \affiliation{
    QTF Centre of Excellence, Department of Applied Physics, Aalto University, P.O. Box 15100, FI-00076 AALTO, Finland
    }

\author{K. V. Petrovnin}%
 \affiliation{
  QTF Centre of Excellence, Department of Applied Physics, Aalto University, P.O. Box 15100, FI-00076 AALTO, Finland
  }
  
  \author{V. Vesterinen}%
 \affiliation{QTF Centre of Excellence, VTT Technical Research Centre of Finland Ltd, P.O. Box 1000, FI-02044 VTT, Finland
 } 
  
 \author{S. Hamedani Raja}%
 \affiliation{
  QTF Centre of Excellence, Department of Applied Physics, Aalto University, P.O. Box 15100, FI-00076 AALTO, Finland
  }
  
\author{I. Lilja}%
 \affiliation{
  QTF Centre of Excellence, Department of Applied Physics, Aalto University, P.O. Box 15100, FI-00076 AALTO, Finland
  }
 
 \author{M. Will}%
 \affiliation{
  QTF Centre of Excellence, Department of Applied Physics, Aalto University, P.O. Box 15100, FI-00076 AALTO, Finland
  }
  
  \author{A. Savin}%
 \affiliation{
  QTF Centre of Excellence, Department of Applied Physics, Aalto University, P.O. Box 15100, FI-00076 AALTO, Finland
  }

 \author{S. Simbierowicz}%
 \thanks{Present address: Bluefors Oy, Arinatie 10, 00370 Helsinki, Finland.}
 \affiliation{QTF Centre of Excellence, VTT Technical Research Centre of Finland Ltd, P.O. Box 1000, FI-02044 VTT, Finland
 }

 \author{R. N. Jabdaraghi}%
 \affiliation{QTF Centre of Excellence, VTT Technical Research Centre of Finland Ltd, P.O. Box 1000, FI-02044 VTT, Finland
 } 
 
  \author{J. S. Lehtinen}%
\affiliation{QTF Centre of Excellence, VTT Technical Research Centre of Finland Ltd, P.O. Box 1000, FI-02044 VTT, Finland
 } 
 
  \author{L. Gr\"onberg}%
 \affiliation{QTF Centre of Excellence, VTT Technical Research Centre of Finland Ltd, P.O. Box 1000, FI-02044 VTT, Finland
 } 
 
 \author{J. Hassel}%
 \thanks{Present address: IQM Quantum Computers, Keilaranta 19, 02150 Espoo, Finland.}
 \affiliation{QTF Centre of Excellence, VTT Technical Research Centre of Finland Ltd, P.O. Box 1000, FI-02044 VTT, Finland
 } 
 
 \author{M. P. Prunnila}%
\affiliation{QTF Centre of Excellence, VTT Technical Research Centre of Finland Ltd, P.O. Box 1000, FI-02044 VTT, Finland
 } 
 
  \author{J. Govenius}%
 \affiliation{QTF Centre of Excellence, VTT Technical Research Centre of Finland Ltd, P.O. Box 1000, FI-02044 VTT, Finland
 } 
  
\author{G. S. Paraoanu}%
 \affiliation{
  QTF Centre of Excellence, Department of Applied Physics, Aalto University, P.O. Box 15100, FI-00076 AALTO, Finland
  }
  
 \author{P. J. Hakonen}%
 \email[Correspondence: ] {pertti.hakonen@aalto.fi}
 \affiliation{
  QTF Centre of Excellence, Department of Applied Physics, Aalto University, P.O. Box 15100, FI-00076 AALTO, Finland
  }

\begin{abstract}
Entangled microwave photons form a fundamental resource for quantum information processing and sensing with continuous variables. 
{\color{black} We use a low-loss Josephson metamaterial comprising superconducting, non-linear, asymmetric inductive elements to generate frequency-entangled photons from vacuum fluctuations at a rate of 2 giga entangled bits per second spanning over 4\,GHz bandwidth.}
The device is operated as a traveling wave parametric amplifier under Kerr-relieving biasing conditions.
Furthermore, we realize the first successfully demonstration of single-mode squeezing in such devices -- $3.1\pm0.7$\,dB below the zero-point level at half of modulation frequency.
\end{abstract}
\maketitle

The generation of quantum resources --  most notably quantum entanglement -- is an essential task for the new emerging industry employing quantum technologies.
While entanglement in discrete variable represents the standard approach for quantum computing, continuous variable (CV) entanglement between microwave photons is a cornerstone for more robust quantum computing \cite{Brecht2016, Magnard2020, Burkhart2020}, sensing \cite{Lloyd2008, Barzanjeh2015}, and communication \cite{Pogorzalek2019, DiCandia2020, Fedorov2021} schemes.
However, despite extensive technical developments over the past decades, the generation of continuous-variable entanglement lacks efficiency in the microwave range.
Thus, it is of great interest to develop new ways to generate robust, high-quality entangled states reliably and at large rate. 

The recent progress in quantum optics using microwaves has stimulated widespread interest in parametric Josephson devices that can serve as both quantum-limited amplifiers \cite{Yurke1988,Yamamoto2008,Hatridge2011,Lahteenmaki2012,Mutus2013,Lahteenmaki2014,Mutus2014, PhysRevB.89.214517, Roy2015,Jebari2018,Elo2019} and quantum resource generators \cite{Eichler2011, Wilson2011, Zhong2013, Lhteenmki2016, Grimsmo2017, PhysRevLett.124.140503}.
Such systems, when modulated by an external pump, convert vacuum fluctuations into real photons, creating squeezing between modes at different frequencies symmetric with respect to half of the frequency of the pump -- 3-wave mixing (3WM), or to the frequency of the pump -- 4-wave mixing (4WM) \cite{Lhteenmki2016}.
One of the most promising devices of this kind is travelling wave parametric amplifier (TWPA) that can operate on a several gigahertz bandwidth \cite{Macklin2015, White2015, Zorin2016, Krinner2019} providing the ability to generate broadband quantum correlations, in contrast to conventional Josephson parametric amplifiers (JPAs) featuring at least an order of magnitude narrower band. 
These broadband characteristics would in principle allow for the operation of a high number of entangled spectrum modes forming a large quantum network \cite{Cai2017}, which can be used for advanced information processing in the microwave range.
However, the performance of real TWPA devices is limited by high loss level, impedance mismatching, and lack of control over nonlinearities \cite{Esposito2021}.

In this work we employ a TWPA where dissipation has been strongly reduced by optimization of the amplification medium. The device allows for adjustable 3-wave (3WM) and 4-wave mixing (4WM) processes. This TWPA has been designed and fabricated at VTT \cite{patent}. 
Using this metamaterial in a Kerr-freed 3-wave mixing mode, we  report the first successful demonstration of vacuum-induced generation of high-quality CV entanglement between frequency-spaced microwaves, reflected also in significant single-mode squeezing.\\

As the TWPA device features a chain of Josephson junction (JJ) based elements for the realization of a non-linear inductance, it is instructive to study the effective potential energy of a single element~\cite{Frattini2017}
\begin{equation}
    U(\varphi) = E_J\left[ c_2(\Phi)\varphi^2 + c_3(\Phi)\varphi^3 + c_4(\Phi)\varphi^4 + \dots\right],
    \label{eq:potential}
\end{equation}
where $E_J$ is the Josephson energy of one of the constituent junctions, $\varphi$ is the superconducting phase difference across the element, and $\Phi$ denotes the external magnetic flux.
Different kinds of Josephson element topologies result in different expansion coefficients $c_n$: the $c_2$ term relates to the critical current and the linear part of Josephson inductance, the $c_3$ term relates to 3WM and the $c_4$ term relates to 4WM, which is also known as the Kerr non-linearity \cite{Yurke2006, Krupko2018}. 
Several topologies have been experimentally proven to work in TWPAs. 
The simplest one is a single JJ \cite{Macklin2015, White2015} $\{c_2,c_3,c_4\}=\{1,0,-\frac{1}{12}\}$, the natural extension of which is a symmetrical superconducting quantum interference device (SQUID)~\cite{Planat2020}. 
To achieve 3WM via a non-zero $c_3$, we choose Josephson elements known as superconducting nonlinear asymmetric inductive elements (SNAILs) with one small, $\alpha E_J$, junction on one side and $n=2$ large, $E_J$, junctions on the other side \cite{Frattini2017}. 
The asymmetry ratio $\alpha<0.5$ ensures that the element exhibits only a single potential minimum.

One critical aspect of TWPAs is the need for phase matching \cite{Agrawal2013}, which was achieved in 4WM TWPAs using resonant matching~\cite{Planat2020, Malnou2021}, or changing the sign of the Kerr non-linearity \cite{Ranadive2021, Bell2015}.
However, the presence of a strong pump at the center of the gain band in 4WM TWPAs is a potential source of back-action.
In practice, the quantum efficiency and output field squeezing are also limited by the Kerr-type non-linearity \cite{Boutin2017}.
Besides that, the generation of higher order harmonics products, increasing the losses in the TWPA \cite{Peng2021}, could lead to substantial degradation of the entanglement and squeezing performance.
\begin{figure}[h]
    \noindent\centering{
    \includegraphics[width=80mm]{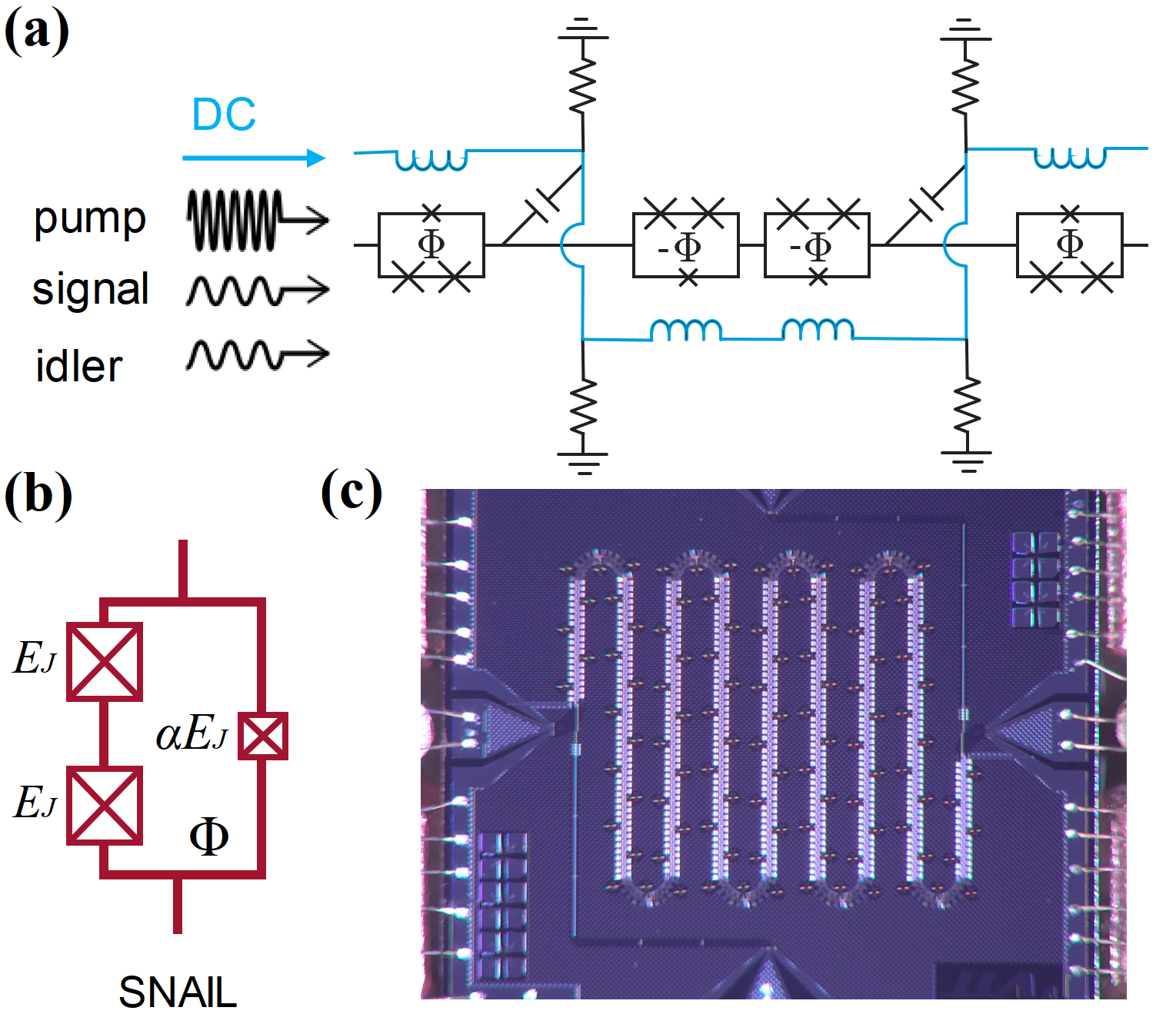}
    }
    \caption{
    (a) Coplanar waveguide transmission line where the center conductor is composed of Josephson elements called Superconducting Nonlinear Asymmetric Inductive elements (in black).
    The DC tuning lines (in blue) are used to change the effective inductance of the TWPA.
    The pump and the signal are combined at the same input port.
    (b) The scheme of SNAIL that provides control over mixing processes via external magnetic flux.
    (c) Optical photograph of the bonded device with chip size $5\times5$ mm$^2$.
    The device features four ports: left and right ports are used for the microwaves input and output and top and bottom additional ports are used for the DC input and output.
    }
    \label{sample}
\end{figure}

Addressing these issues, our TWPA is a coplanar waveguide (CPW) transmission line where a SNAIL array center conductor simultaneously suppresses the Kerr mixing term and ensures $50\,\Omega$ impedance matching, while retaining a sufficient degree of 3WM.
In contrast to Ref.~\onlinecite{Bell2015} where an alternating polarity of SNAIL-like elements suppresses 3WM, our SNAILs (Fig.\,\ref{sample}b) have the same dipolar orientation throughout.
We realize homogeneous flux biasing for all the 1,632 SNAILs, and negligible off-chip fringing fields, using a magnetic flux bias line placed on chip.
The details of the unit cell design that is shown in Fig.\,\ref{sample}a are presented in Refs.\,\onlinecite{patent, SI}. 
Resistors keep the flux bias line (depicted in blue in Fig.\,\ref{sample}a) grounded at microwave frequencies, thereby acting as L/R lowpass filter and blocking microwave propagation along it. 
The scheme also contributes towards keeping the two sides of the CPW ground at the same potential. 
At the same time the resistors prevent leakage of the static current and the formation of parasitic loops of quantized magnetic flux. 
The resistance value is chosen to be much lower than the reactance of the unit cell shunt capacitor at signal, idler and pump frequencies -- 1200, 1200 and 600 times respectively.
The resistors thus contribute to the effective loss tangent of the shunt (quantified below), which, however, remains dominated by dielectric loss.

The device (Fig.\,\ref{sample}c) was fabricated with a side-wall passivated spacer niobium junction technology detailed in Refs.~\onlinecite{Gr_nberg_2017, Simbierowicz2021}. 
At zero flux bias, the characteristic impedance of the TWPA is far below the optimal value of $50\,\Omega$ and the SNAIL non-linearity does not allow for 3WM.
Ramping up the bias into the regime near $\Phi=0.4\,\Phi_0$, where $\Phi_0$ is magnetic flux quantum, improves the impedance matching, and the SNAILs exhibit 3WM combined with only weak 4WM. 
The weakened 4WM contributes favorably to the phase matching of the amplification process and to non-classical state generation. 

In a linearly dispersive TWPA the 3WM would strongly and unidirectionally convert pump power into higher harmonics~\cite{Parameswaran2002}. 
At angular frequencies higher than the pump ($\omega_p$), our TWPA has non-linearly dispersive features primarily from the combination of Josephson plasma resonance and the periodic loading of the CPW with dispersive capacitors~\cite{Malnou2021, patent2}. 
We control the phase mismatch of the second harmonic generation (SHG)~\cite{PhysRevApplied.14.034058} in particular. 
Instead of blocking SHG altogether we allow a weak and cascaded mixing process where the pump accrues a favorable phase shift when converting from $\omega_p$ to $2\omega_p$ and later back to $\omega_p$~\cite{DeSalvo1992}. 
The phase shift looks effectively like 4WM and contributes to the compensation of the Kerr effect within the device. 
The pump frequency is a degree of freedom that allows us explore different levels of SHG and, consequently, to optimize the pump phase shift~\cite{patent2}.

Our device provides 15.3\,dB of gain over 3\,GHz bandwidth with the noise temperature being close to the single-photon quantum limit.
The results on TWPA characterization as a quantum-limited amplifier are presented in Ref.\,\onlinecite{SI}. 
Although our loss tangent of about $0.003$ isn't particularly low, 3WM requires a shorter transmission-line length for the same amount of gain as 4WM~\cite{Zorin2016}.
At our frequency of interest $4.8$\,GHz, the unpumped TWPA shows about $0.6$\,dB of loss~\cite{SI}, which is to be contrasted against $2.4$\,dB and $3.0$\,dB extrapolated from Refs.~\onlinecite{Planat2020, Bultink2018}.
Such a low loss level is the key feature that allows us to generate entangled quantum states and achieve squeezing.  \\

\begin{figure}
    \noindent\centering{
    \includegraphics[width=80mm]{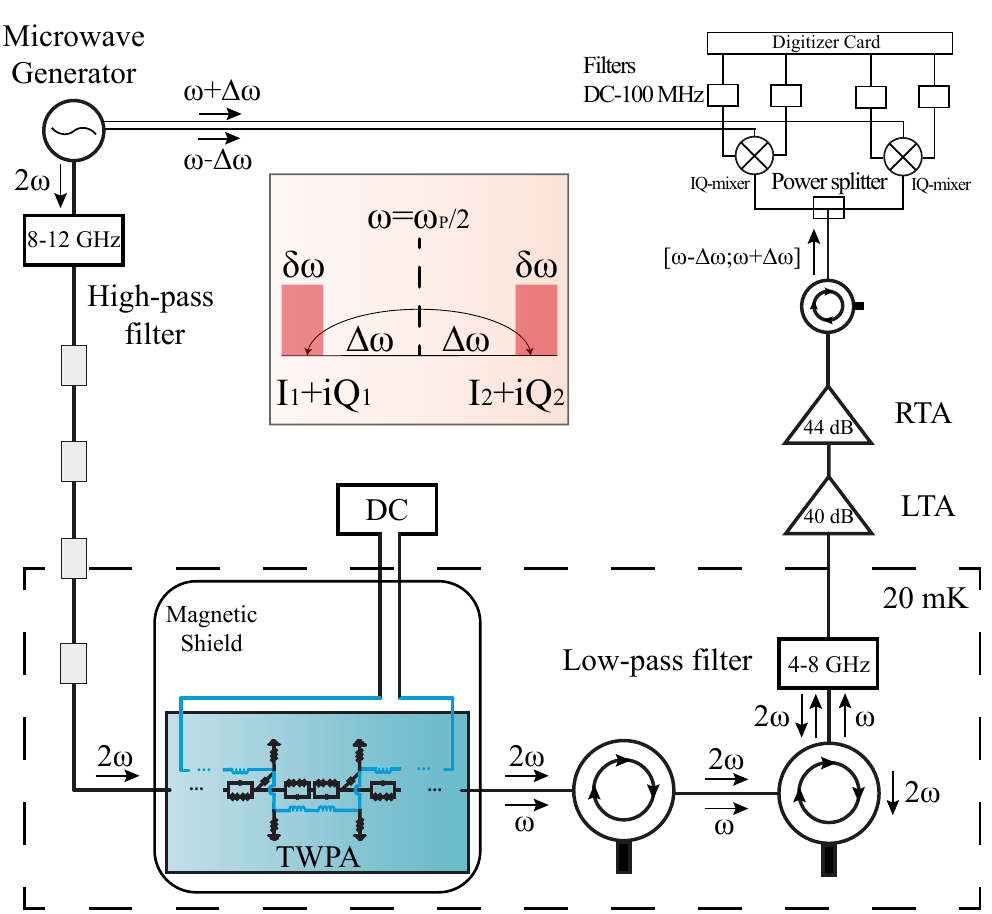}
    }
    \caption{Measurement scheme for entanglement generation and verification.
    The microwave pump at frequency $2\omega$ enters the TWPA through the chain of filters and attenuators and triggers the generation of correlated photons.
    The TWPA is biased with a magnetic flux via DC line.
    The pump signal is filtered out with a band-pass filter, while the 4-8\,GHz microwaves are amplified with a chain of low-temperature and room-temperature amplifiers, LTA and RTA, respectively.
    {\color{black} The output signal in $[\omega_p/2-\Delta\omega;\omega_p/2+\Delta\omega]$ range is split into two branches using a power splitter. 
    The two branches are demodulated with separate local oscillators to symmetric frequency bands. 
    Outputs of the IQ mixers are then filtered to avoid aliasing effects, after which the signals are digitized with an Alazar digitizer card.}
    The scheme of the generation of frequency-entangled microwaves is presented in pink embedding.
    Two modes $\Delta_L$ and $\Delta_R$ with $\delta\omega$ bandwidth are shifted by $\Delta\omega$ from the $\omega_p/2$ frequency.
    }
    \label{scheme}
\end{figure}

The experimental scheme for TWPA-based entanglement generation is presented in Fig.\,\ref{scheme}.
The pump signal $\omega_p=2\omega$ enters the TWPA at 20\,mK temperature through the chain of attenuators and triggers the generation of correlated photons in the frequency bands $\Delta_L = [\omega-\Delta\omega-\delta\omega : \omega-\Delta\omega]$ and $\Delta_R =[\omega+\Delta\omega:\omega+\Delta\omega+\delta\omega]$ generated from vacuum fluctuations \cite{Lhteenmki2016}.
Here, $\Delta\omega$ is band detuning from frequency $\omega$, and $\delta\omega$ is the bandwidth that is set manually and can be specified according to the goals.
{\color{black} In our experiment, $\omega=\omega_p/2=2\pi\times4.8$\,GHz, $\delta\omega=2\pi\times0.5$\,MHz and band separation $2\Delta\omega$ is varied from 0, the case of  minimal mode separation, up to $2\pi\times4$\,GHz, the maximum mode separation allowed by the bandwidth of our experimental setup.}
Both $\omega$ and $2\omega$ signals exit the TWPA, but only a filtered subband of 4-8\,GHz microwaves are amplified using the HEMT and room-temperature amplifiers.
{\color{black} 
We record signals using dual-readout configuration, which simultaneously demodulates the bands with two separate IQ mixers \cite{Zhong2013} that are then digitized with a multi-channel Alazar ATS9440 digitizer.}

We characterize the entanglement in $N$-mode output quantum state using the covariance formalism.
According to the theory of parametric amplifiers, all output states are Gaussian \cite{Braunstein2005, Adesso2014}.
We verify this assumption by measuring the skewness and kurtosis that prove the Gaussian nature of the signals \cite{SI}.
The state is fully characterized by the covariance matrix $\mathbf{V}$ of the $I$ and $Q$ voltage quadratures of the propagating microwave modes $1, ..., N$:
\begin{align}
    K = (I_1, Q_1, \dots, I_N, Q_N)^T,\nonumber \\
    V_{ij} = \langle K_{i}K_{j}+K_jK_i \rangle/2.
\end{align}

Before the covariance matrix analysis, we calibrate the quadratures in order to be able to compare the variances and covariances to the vacuum noise level, which is crucial for the entanglement verification.
We convert quadrature voltages $I$ and $Q$ to the scaled quadratures $\mathcal{I}$ and $\mathcal{Q}$ in the following way:
\begin{align}
    \langle \mathcal{I}^2_i \rangle = \frac{\langle{I^2_i}\rangle_{ON}-\langle{I^2_i}\rangle_{OFF}}{\mathcal{N}}+\mathbb{1}, \nonumber\\
    \langle \mathcal{I}_i \mathcal{I}_j \rangle = \frac{\langle{I_i\,I_j}\rangle_{ON}}{\mathcal{N}},
\end{align}
with the same transformation applied to $\mathcal{Q}$.
Here, the normalization coefficient $\mathcal{N}$ is given by
\begin{equation}
    \mathcal{N}=\frac{G_{sys} Z_0 h f BW}{4\eta},
\end{equation}
where $G_{sys}$ is measured system gain, $Z_0=50\,\Omega$, $h$ is Planck's constant, $f$ is operating frequency, $BW$ is the measurement bandwidth and $\eta$ represents internal losses of the TWPA.
We rescale the vacuum fluctuations to unity for analysis.
Details on system gain calibration and rigorous analysis of quantum modes propagating in the TWPA within distributed gain and loss model are presented in Ref.\,\onlinecite{SI}.

\begin{figure}
    \noindent\centering{
    \includegraphics[width=80mm]{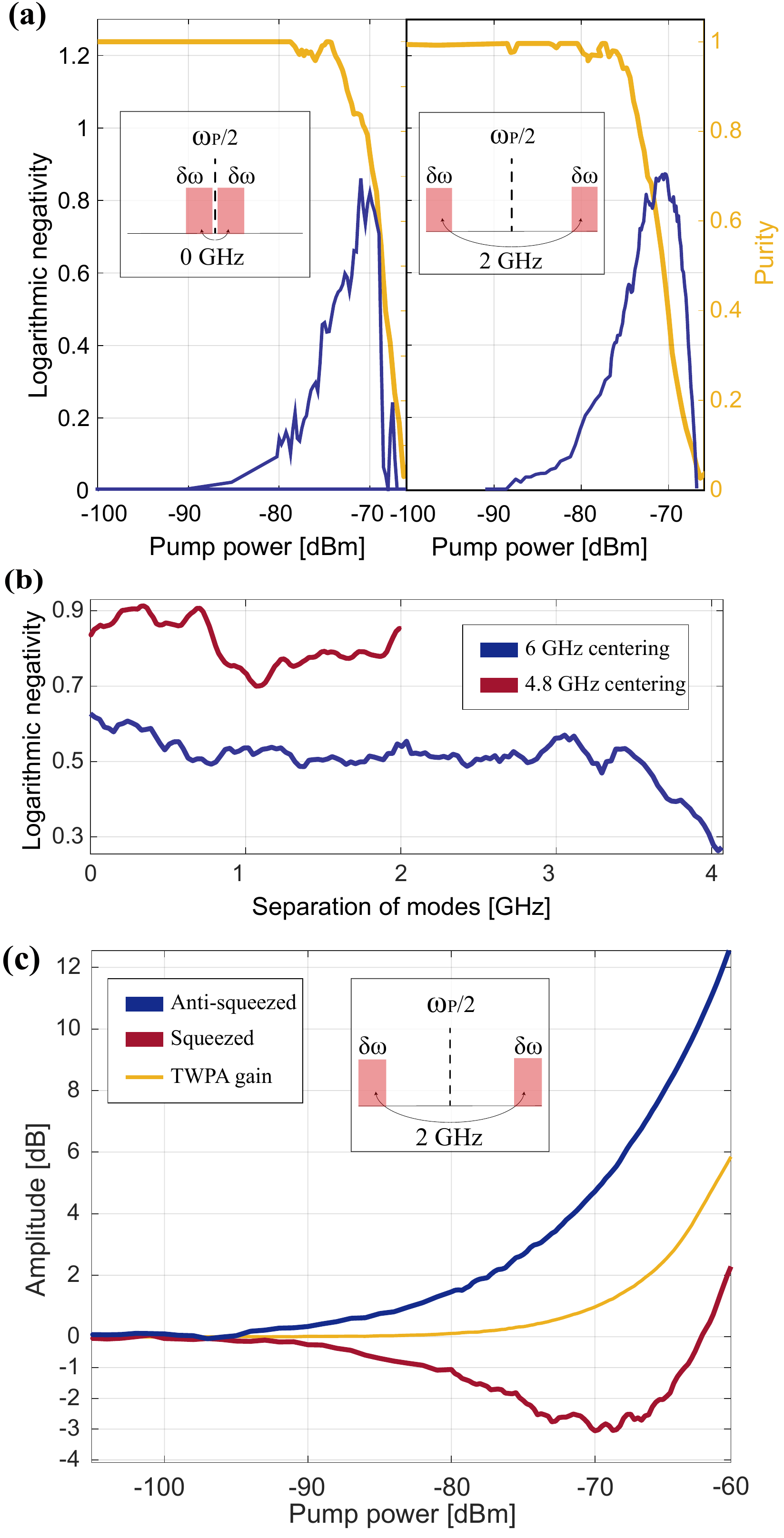}
    }
    \caption{
    (a) Logarithmic negativity $E$ and purity as function of the pump power {\color{black} with $\omega_p=2\pi\times9.6$\,GHz.}
    Non-zero negativity indicates the presence of entanglement.
    Negativity for non-spaced frequency bands is depicted in the left frame, while negativity for {\color{black} 2\,GHz mode separation is shown in the right frame.}
    The purity is shown in yellow: as purity fades, the entanglement disappears. 
    The insets show the frequency mode definition.
    {\color{black} (b) Maximum logarithmic negativity as function of the separation of modes with optimised pump power for two pump frequency $\omega_p=2\pi\times11.98$\,GHz (in blue) and $\omega_p=2\pi\times9.6$\,GHz (in red).
    While pumping at $\omega_p=2\pi\times9.6$\,GHz allows for the higher logarithmic negativity, the center of the operating band is not in the center of experimental band, which limits the separation of modes.
    In case of $\omega_p=2\pi\times11.98$\,GHz, the entanglement is preserved over the whole 4\,GHz bandwidth.
    (c) Two-mode squeezing for different pump powers in case of $\omega_p=2\pi\times9.6$\,GHz.
    Amplified $S_-$ quadrature is depicted in blue, squeezed $S_+$ quadrature is depicted in red.
    The gain of the TWPA is shown in yellow. } }
    \label{quantum_state}
\end{figure}

Our entanglement analysis employs the Peres-Horodecki criterion of positivity under partial transpose (PPT) \cite{Peres1996, Horodecki1997} based on symplectic transformation of the covariance matrix \cite{Simon2000}.
In a further analysis, we use the logarithmic negativity \cite{Vidal2002}, $E = \mbox{max}[-\log_2{\nu_{min}}, 0]$, that constitutes an upper bound to the distillable entanglement of the state and corresponds to the entanglement cost under PPT preserving operations \cite{Audenaert2003}.
{\color{black} Here $\nu_{min}$ is a minimum symplectic eigenvalue.}
Considering the scaled quadratures which correspond to vacuum fluctuations equal to $1$ (see Ref.\,\onlinecite{SI}), $E$ is non zero only if $\nu_{min}<1$, and it quantifies the quantum information capacity of the entangled state.
We want to emphasize that the low loss level is crucial for the entanglement generation since the internal loss parameter $\eta$ defines the normalization coefficient $\mathcal{N}$, which, in turn, affects the minimum symplectic eigenvalue.\\

{\color{black} We characterize our results on two-mode ($N=2$) entanglement in terms of logarithmic ($\log_2$) negativity.}
Fig.\,\ref{quantum_state}a displays $E$ as function of pump power for two different separations of $\Delta_L$ and $\Delta_R$ modes.
Mixers used for the downmixing of output signals in the measurement scheme introduce correlations that have classical nature with $E_0\approx0.1$.
Since $E$ is an additive quantity, we subtract $E_0$ measured at $-110$\,dBm from the data resulting in $E=0$ at low powers that indicates an uncorrelated ground state of the metamaterial.
{\color{black} With increasing of pump power, the logarithmic negativity reaches $0.85\pm0.20$ for non-spaced modes at $-69$\,dBm power as shown in the left frame in Fig.\,\ref{quantum_state}a.
For modes with 2\,GHz spacing, we observe similar behaviour with the same maximum logarithmic negativity indicated in the right frame in Fig.\,\ref{quantum_state}a.}
With further increase in pump power, the entanglement disappears since high drive power introduces detrimental higher order couplings and wave-mixing processes \cite{SandboChang2018}.
{\color{black} Similar behaviour has been observed in transmission lines terminated by a SQUID \cite{PhysRevLett.124.140503} and lumped-element JPAs \cite{Boutin2017}.}
In order to prove the nonideality of the state at higher pump powers, we show the decline of purity, calculated for Gaussian states \cite{Paris2003}, in Fig.\,\ref{quantum_state}a for both spaced and non-spaced modes.

{\color{black}
To demonstrate the broadband entanglement generation we fix the pump frequency and analyse the maximum $E$ as a function of separation of modes, see Fig.\,\ref{quantum_state}b, and verify the entanglement generation over 2\,GHz bandwidth, which is limited by the experimental scheme.
In order to generate entanglement over the whole 4\,GHz bandwidth, we change the pumping frequency $\omega=\omega_p/2=2\pi\times5.99$\,GHz so that the half of the pump was at the center of the measurement band. 
Such a change in pump frequency degrades impedance matching, resulting in lower logarithmic negativity value, as demonstrated in Fig.\,\ref{quantum_state}b.
The internal losses and the system gain is measured and calibrated over the whole $4-8$\,GHz bandwidth.
As can be seen, the logarithmic negativity is nearly constant and starts to decrease only near the bandwidth limit of the experimental setup.
}

To illustrate the excellent quality of our metamaterial, we demonstrate the generation of vacuum-induced squeezed states.
{\color{black} For this experiment, we consider two modes with 2\,GHz spacing centered at $\omega_p/2$.}
The parametric amplifier operates in the degenerate mode at this point.
Measuring and calibrating $I$ and $Q$ quadratures in the same way as was for two-mode entanglement verification, we introduce $S_+=\langle \mathcal{I}^2 \rangle$ and $S_-=\langle \mathcal{Q}^2 \rangle$ amplitudes that correspond to squeezed and amplified quadratures with respect to the vacuum level.
These amplitudes are shown in Fig.\,\ref{quantum_state}c as function of the pump power.
{\color{black} The maximum squeezing of $3.1\pm0.7$\,dB is achieved at $-69$\,dBm pump power.}
Higher powers introduce non-linearities that distort the quantum state and reduce the squeezing, similarly to what occurs for resonant Josephson parametric amplifiers \cite{Boutin2017}.
{\color{black} The same level of squeezing was observed for the single-mode experiment and presented in \cite{SI}.}

In order to quantify our TWPA as a generator of entangled photon flux, we define an entanglement generation rate
\begin{equation}
    R_E = 2 \langle \mathcal{I}_1^2 \rangle\,E_F (\Delta\omega+\delta\omega)
\end{equation}
that quantifies the production rate of available quantum information.
Here, $\langle \mathcal{I}_1^2 \rangle$ is the estimation on the photon flux intensity (\,$\sim 1/(\mathrm{s\,Hz})$\,), $E_F$ is the entropy of formation in ebits \cite{Bennett1996} and $2(\Delta\omega+\delta\omega)$ is the effective bandwidth of the frequency span in which photons are generated from vacuum fluctuations.
The entropy of formation quantifies how many Bell states are needed to prepare the given state using local operations and classical communications, and it is calculated from the minimum symplectic eigenvalue $\nu_{min}$ detailed in \cite{SI}.
{\color{black} 
We find that our device provides at least $2$\,Gebits/s (giga entangled bits per second) over 4\,GHz, being limited 
by the working band of the experimental setup. 
Outside  of the $4-8$\,GHz band, the signal is cut off by circulators, bandpass filters and the bandwidth of the HEMT amplifier, making the entanglement verification impossible at these frequencies.}
A similar JPA-based entanglement generation scheme featuring 5\,dB of squeezing ($\nu_{min}=0.33$) and 5\,MHz bandwidth \cite{Zhong2013} can provide only 6\,Mebits/s -- {\color{black} three orders of magnitude lower value}.\\

In summary, we have demonstrated a broadband SNAIL Josephson travelling wave parametric amplifier for generation of non-classical states.
{\color{black}
Owing to low losses in our 3WM TWPA design, we have been able to generate entangled microwave signals over 4\,GHz bandwidth, and with $E=0.85\pm 0.20$ maximum logarithmic negativity with optimized pumping.
For single frequency mode of operation, we reached $3.1\pm0.7$\,dB of quantum squeezing.
Higher entanglement generation rate and squeezing can be achieved by further improving the experimental scheme and the TWPA, mainly impedance matching, reducing dielectric losses and dissipation by resistance, and suppressing the Kerr nonlinearity.}\\

Finally, the results presented in this work hold high promise in meeting the challenges of broadband quantum information processing with continuous-variable (CV) states. 
{\color{black} Indeed, the Kerr effect can be employed in realizing universal quantum gates for CV quantum computing \cite{Lloyd1999,Bartlett2002,Yanagimoto2020}.}
In Josephson microwave circuits, Kerr effects are much stronger than in optical systems, and, with the excellent engineering of nonlinearities demonstrated in this work, these proposals may turn out to be realizable.
{\color{black} As we demonstrated, the broadband features of the TWPA allow operation over few gigahertz bandwidth}, and in combination with the multiple pumping scheme \cite{Petrovnin2021} pave the way towards generation of frequency-spaced multimode entanglement.
Multimode schemes can be employed for various quantum applications, such as CV computing with cluster states \cite{Hillmann2020}, secure and robust communications \cite{Samsonov2020}, distributed quantum-limited sensing \cite{Guo2019} and search for dark matter \cite{Backes2021}.\\

{\color{black} During the peer-review process we became aware of the works on squeezing and entanglement generation in TWPAs \cite{Esposito2021twpa, Qiu2022}.}\\

\begin{acknowledgments}
We thank Alpo Ahonen, Paula Holmlund and Harri Pohjonen for technical assistance and Terra Quantum AG for scientific support. KVP has been funded by the European Union's Horizon 2020 research and innovation programme under grant agreement no. 862644 (FET-Open project: Quantum readout techniques and technologies, QUARTET). 
The work at VTT has been funded from the EU Flagship on Quantum Technology Grant No. H2020-FETFLAG-2018-03 Project Nos. 820363 OpenSuperQ and 820505 QMiCS. 
This project has received funding from the European Union’s Horizon 2020 research and innovation programme under grant agreement no. 824109 (supporting MRP, IL, MW, and AS), no. 862644 (FET-Open project QUARTET), and ERC grant agreement no. 670743 (QuDeT).
PJH and VV acknowledge financial support from the Academy of Finland through grants nos. 314448 and 321700, respectively. The  work  of  SHR  and PJH was  supported  by  MATINE research grant. GSP and KVP would like to thank Saab for scientific collaboration under a research agreement with Aalto University.
This work has been done under the “Finnish Center of Excellence in Quantum Technology QTF” of the Academy of Finland, project nos. 312059, 312294, 312295, 336810, and 312296.
\end{acknowledgments}



\onecolumngrid
\newpage
\begin{center}
{\LARGE  Supplemental material for Broadband continuous variable entanglement generation using Kerr-free Josephson metamaterial}\\
\end{center}

\section{Details of TWPA design and fabrication}
\begin{figure}[ht]
    \noindent\centering{
    \includegraphics[width=86mm]{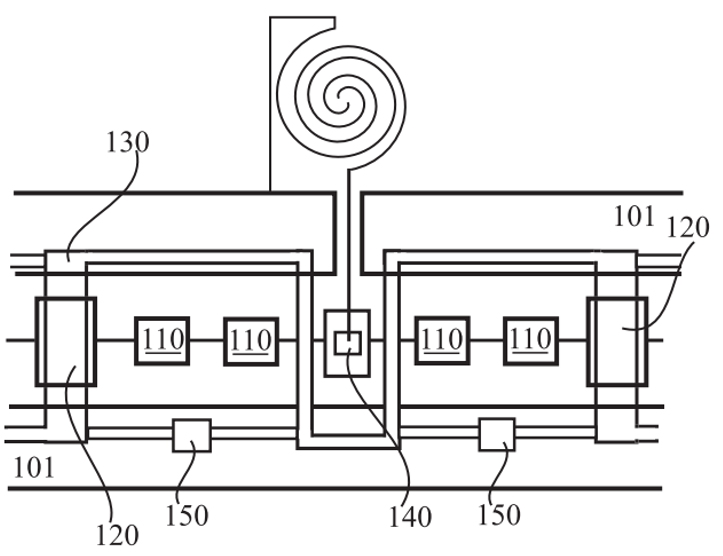}
    }
    \caption{ 
    A top-view cartoon of the TWPA layout on the chip.
    }
    \label{unitcell}
\end{figure}

The TWPA is realized with VTT's multilayer superconductive fabrication process~\cite{Gr_nberg_2017, Simbierowicz2021}, where the metal and insulator thicknesses vary between 50--200~nm. The fabrication on a high-resistivity silicon wafer begins with the deposition and patterning of a Nb/Al-AlO$_x$/Nb trilayer. We use strips of the trilayer in the SNAILs (110) in Fig.~\ref{unitcell}. Whenever the subsequent main wiring layer of niobium crosses with a strip, a Josephson junction will form at the cross-over. The ground planes (101) of the coplanar waveguide (CPW) are made of the main wiring layer. The main wiring layer also has dispersive capacitors (open-ended transmission-line spiral resonators \cite{Yan2021, patent2} with one side galvanically connected to the CPW center conductor (140) and the other side to the ground) as well as the bottom electrodes of parallel-plate capacitors (120). Next, we uniformly cover the device with atomic layer deposited aluminum oxide insulator. We make contact holes to the insulator at the forthcoming sites of resistors (150), for instance. We cover the insulator with a secondary niobium wiring layer that is used for the magnetic flux bias line (130). The flux line has a meander shape and it repeatedly crosses from one side of the CPW to another. Notably, the flux line also serves as the top electrode of the parallel-plate capacitors (120)~\cite{patent}. The final fabrication step is the deposition and patterning of a normal-metal resistive layer. The resistors (150) connect the flux line on the secondary wiring layer to the ground (101) on the main wiring layer.

{\color{black} For details on managing the higher harmonic generation of the pump frequency, we refer to Ref.~\cite{patent2}. In brief, we arrange a TWPA dispersion curve that is superlinear at the second harmonic generation (SHG) frequency of the pump. The SHG, which is inherent to 3WM, is phase mismatched and the frequency-doubled microwaves are downconverted with a shifted phase. This refractive effect, which resembles the Kerr effect in 4WM, is well known in the field of non-linear optics~\cite{DeSalvo1992}. 
The primary origin of the superlinear dispersion is a stopband near the SHG frequency, induced by the periodic loading of the TWPA with the spiral resonators. 
The first stopband from the loading occurs when the distance between the spirals corresponds to half a wavelength~\cite{Malnou2021}. 
The dispersive effects of the spiral resonator eigenmode and the Josephson plasma frequency are responsible for the phase mismatching of any higher-order non-linear processes above the SHG frequency.}

\section{Control of the wave mixing process in TWPA}

To address the non-linearity of the device, we first derive the potential energy of a single SNAIL. We denote the phase of the smaller Josephson junction  by $\varphi_s$, and the total phase of the SNAIL (traversing around the loop) by $\varphi_{ext}=2\pi \Phi/\Phi_0$. With uniform phase drop across the two large junctions, transport current of the SNAIL reads $I=\alpha I_c \sin{\varphi_s}-I_c\sin{(\varphi_{ext}-\varphi_s)/2}$, where $I_c$ is the critical current. Having the voltage-phase relation $V=\Phi_0/2\pi~\Dot{\varphi}$, one obtains the potential energy of the SNAIL by
\begin{equation}
    U_S(\varphi_s)=E_J\int \Big(\alpha I_c \sin{\varphi_s}-I_c\sin{\frac{\varphi_{ext}-\varphi_s}{2}}\Big)\Dot{\varphi_s}=-\alpha E_J \cos{\varphi_s}-2E_J\cos{\frac{\varphi_{ext}-\varphi_s}{2}},
\end{equation}
with $E_J=\Phi_0 I_c/2\pi$. Assume now that $U_S$ has a minimum at a phase denoted by $\varphi_{min}$. We Taylor-expand the potential energy around this point, such that with $\Tilde{\varphi}=\varphi_s-\varphi_{min}$ we have
\begin{equation}
U_S(\Tilde{\varphi})/E_J=c_1\Tilde{\varphi}+c_2\Tilde{\varphi}^2+c_3\Tilde{\varphi}^3+c_4\Tilde{\varphi}^4+\dots,
\end{equation}
where $c_n=(1/n!E_J)d^n U_S(\varphi_s)/d\varphi_s ^n\vert \varphi_{min}$. Sitting at the minimum of $U_S$ means that $c_1=0$, i.e., the transport current of SNAIL is zero. The coefficients $c_n(\varphi_{ext})$ and the Josephson inductance $L_J=\Phi_0/2\pi I_c$ completely determine the behaviour of a SNAIL. However, having an array of SNAILs embedded in a transmission line is more involved and requires careful consideration of the non-linear current conservation. Here, we follow the results presented in \cite{Frattini2018} to address this issue and to estimate the non-linearity of our device, which determines 3WM and 4WM. Consider a simple lumped-element model in which an array of $M$ SNAILs are placed in series with a linear inductance $L$ and capacitance $C$. Taking $\varphi$ as the canonical phase coordinate for the mode, the potential energy of the whole circuit reads
\begin{equation}
    U_T(\varphi,\varphi_s)=MU_S(\varphi_s)+\frac{1}{2L}\varphi_0^2(\varphi-M\varphi_s)^2,
\end{equation}
where $\varphi_0$ is the reduced flux quantum. As explained in \cite{Frattini2018}, the SNAIL phase $\varphi_s$ is no longer an independent coordinate and needs to be treated as a function of the phase coordinate $\varphi$. This can be achieved by accounting for the non-linear current conservation for the node between the array of SNAILs and the linear inductance:
\begin{equation}
 \alpha I_c \sin{\varphi_s} + I_c \sin{\frac{\varphi_s-\varphi_{ext}}{2}}+I_c \frac{L_J}{L}(M\varphi_s-\varphi)=0.
\end{equation}
Taking into account $\varphi_s[\varphi]$, Taylor expansion of the total potential energy $U_T$ around a minimum determined by $\Bar{\varphi}_{min}$ yields the coefficients $\Tilde{c}_n$ which are related to the corresponding coefficients for a single SNAIL \cite{Frattini2018}:
\begin{equation}
    \Tilde{c}_2=\frac{p}{M}c_2,~\Tilde{c}_3=\frac{p^3}{M^2}c_3,~\Tilde{c}_4=\frac{p^4}{M^3}\Big(c_4-\frac{9c_3^2}{4c_2}(1-p)\Big),
\end{equation}
with $p=\frac{ML_J/L}{2c_2+ML_J/L}$.
4WM is connected to $\Tilde{c}_4$ (Kerr term). According to the expressions above, this term is dependent on $c_3^2$ in addition to $c_4$. Therefore, a vanishing $c_4$  does not appear to be sufficient for a Kerr-free situation.

\begin{figure}[ht]
    \noindent\centering{
    \includegraphics[width=90mm]{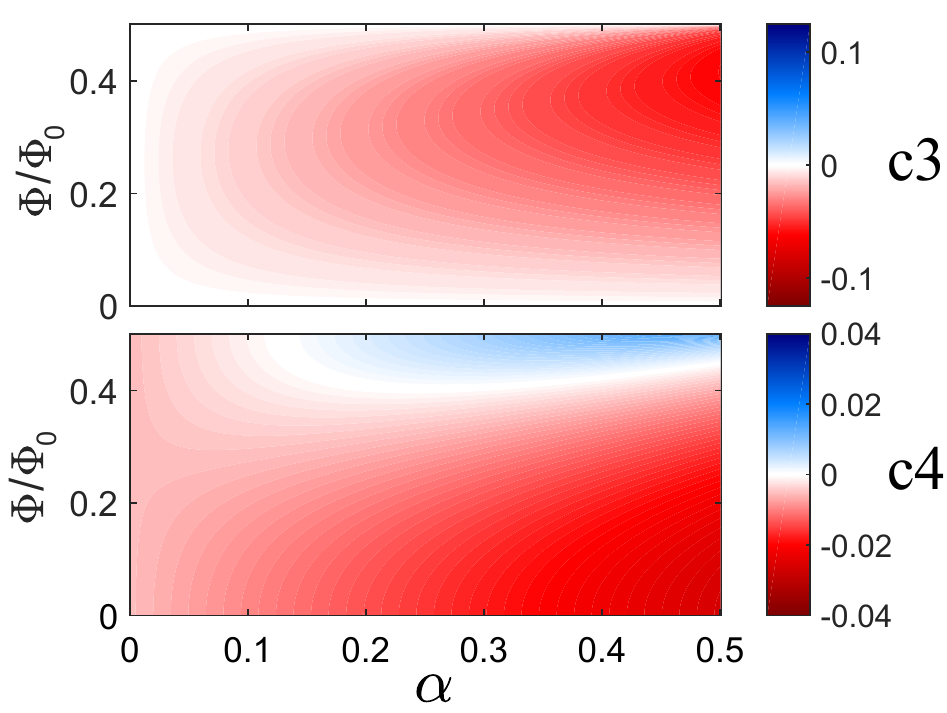}
    }
    \caption{ 
    The dependence of $c_3$ and $c_4$ coefficients of a SNAIL on the applied magnetic flux and SNAIL asymmetry $\alpha$. 
    }
    \label{nonlinearities}
\end{figure}

The results of the simple lumped-element model may not be directly applicable to a TWPA as such, but we expect that basically similar observations will hold for coupled-mode equations~\cite{Zorin2016, Malnou2021} describing the waves propagating in the TWPA.  All through the TWPA operation, it is important to keep in mind  that the effective Kerr coefficient depends not only on $c_4$ but also on a term proportional to $c_3^2$. As a consequence, we take both $c_3$ and $c_4$ into account for optimizing the device. The coefficients $c_3$ and $c_4$ are shown in Fig.\,\ref{nonlinearities} for various flux and SNAIL asymmetry values. 3WM requires a magnetic flux $0<|\Phi|<0.5\Phi_0$, where $\Phi_0$ is the magnetic flux quantum. Note that $c_3(-\Phi) =-c_3(\Phi)$, while $c_4(-\Phi)=c_4(\Phi)$.

\section{Measurement setup}
The TWPA was cooled down in a dry dilution refrigerator BlueFors LD400 with 10\,mK base temperature.
The non-magnetic package for TWPA was made of a copper base, an aluminum lid, and brass connectors with two SMP ports for microwaves and two MMCX ports for the on-chip flux line.
The pump signal was generated by ANAPICO APMS12G 4-channel signal generator with -150\,dBc/Hz phase noise at 100\,kHz.
The pump signal was band-pass limited to 8-12\,GHz range with Mini-Circuits filters to ensure suppression of $\omega$ signal and was 
thermalized by a chain of attenuators, -60\,dB altogether.
The DC signal was generated by Stanford SIM928 Isolated Voltage Source, filtered and routed to flux loops by a superconducting twisted pair cable.
{\color{black} In order to guide the signals to the output, we use Low Noise Factory circulators 4-12\,GHz operating as isolators.
The filtered signal in 4-8\,GHz range is first amplified by a cooled Low Noise Factory LNC4 8C HEMT amplifier (40\,dB of gain and 4\,K noise temperature), connected using a BeCu coaxial cable and SMA connectors to a room-temperature amplifier with 44\,dB gain.
The microwaves are captured with the four-channel 14 bit Alazar ATS9440 digitizer with 125\,MS/s rate.
}

\section{TWPA characterization}
\begin{figure}[ht]
    \noindent\centering{
    \includegraphics[width=100mm]{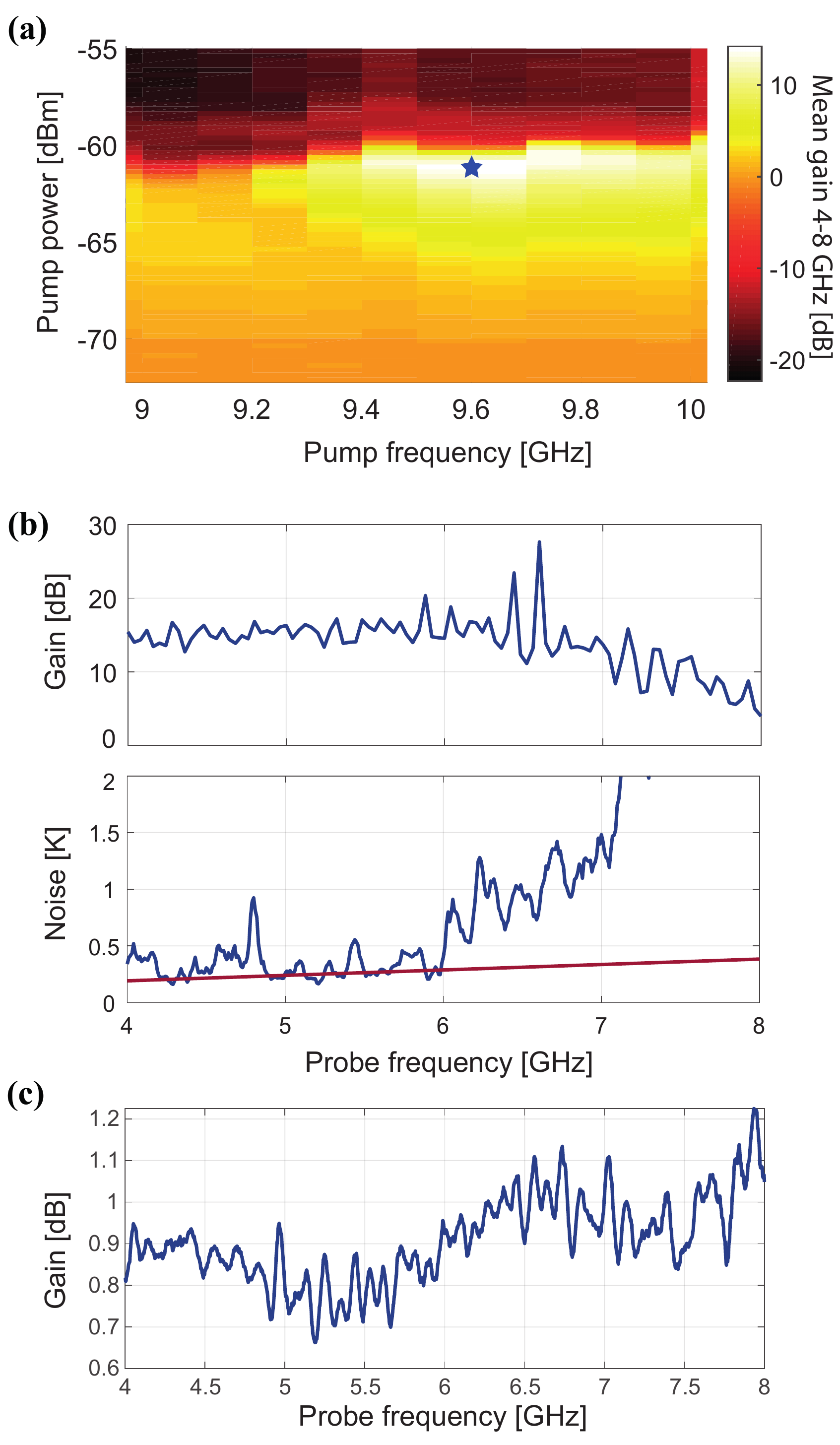}
    }
    \caption{ 
    (a) Mean gain for different pump powers and frequencies measured over 4-8\,GHz range. 
    DC-bias flux is $\Phi=0.38\,\Phi_0$.
    The blue star marks the operation point where gain profile and noise temperature were measured.
    (b) Gain profile and noise temperature of TWPA over 4-8\,GHz range at the operating point marked in frame (a).
    While the TWPA provides 13\,dB of gain on average, the noise temperature (blue line at the bottom) is close to the one-photon quantum limit (red line) $\hbar\omega/k_B$ over 2\,GHz bandwidth, with $k_B$ as the Boltzmann constant.
    {\color{black} (c) Gain \emph{vs.} signal frequency at low pump power which corresponds to the maximum entanglement generation.}
    }
    \label{amplifier}
\end{figure}

First, we characterize our device from the perspective of broadband low-noise parametric amplifier in a dry dilution refrigerator at $T=20$\,mK.
For this purpose, we probe the TWPA with a weak microwave signal (-120\,dBm) and apply a pump through the same port, and record the amplified signal. 
We characterize the amplification performance by measuring the mean gain value normalized against the superconducting transmission line placed in the TWPA's sample holder
over 4--8\,GHz frequency range. 
The best gain performance of the device was observed at a static flux bias of $\Phi=0.38\,\Phi_0$: the mean gain is depicted in Fig.\,\ref{amplifier}a for various pump powers and frequencies.
Choosing the best operating point with the pump frequency 9.6\,GHz and the pump power -63\,dBm (marked with the blue star in Fig.\,\ref{amplifier}a) we evaluate the gain profile depicted in Fig.\,\,\ref{amplifier}b.
The 1-dB compression point indicates the input saturation power of the amplifier is $-100$\,dBm for 20\,dB gain.

The noise performance of the TWPA was characterized using the signal to noise ratio improvement method, giving noise relative to system noise temperature \cite{Roy2015}.
The noise temperature of the device is depicted in Fig.\,\ref{amplifier}b on the bottom.
As can be seen from the figure, the noise is close to the single-photon quantum limit over 2\,GHz bandwidth centered at 5\,GHz, being 350\,mK on average (1.45 photons).
{\color{black} In the case of low pump power, the TWPA features broadband, flat gain profile, see Fig.\,\ref{amplifier}c.
We operate the device in such a regime, because the entanglement is generated at low gain.}

The quantum efficiency \cite{Boutin2017} is then limited by the noise introduced by TWPA. 
Efficiency $\eta_e=1/(1+N)$, where $N$ is the number of added noise photons, is $\eta_e=0.69$ in the case of high TWPA gain. 
It is impossible to characterize the added noise when the gain of TWPA is low in the exploited experimental setup.

\section{Noise calibration}
The scheme for the system gain and noise calibration is presented on Fig.\,3 in the main text.
The calibration was done by measuring the power of the Johnson-Nyquist noise as a function of temperature.
During the calibration we vary the noise power generated by components at 20\,mK stage by changing the physical temperature of the whole stage.
We assume that the output noise of the heated components is fully coupled to the impedance-matched measurement system.
The Johnson-Nyquist noise power is given by $P(T)=BW k_B G_{sys} (T_N+T)$, where $k_B$ is the Boltzmann constant, $BW$ is the measurement bandwidth, which is $1$\,MHz in the experiment, and $G_{sys}$ and $T_{sys}$ are system gain and noise temperature that we aim to estimate.
In order to determine the noise power, we measure the temperature-dependent variance of the output voltage, $\mbox{Var}[V_{IQ}]$, which was captured at different temperatures using an Anritsu MS2830A Signal Analyzer.
The same signal analyzer was used for the entanglement verification.
The resulting noise power is given by $P(T)=\mbox{Var}[V_{IQ}(T)]/Z_0$; here $Z_0$ is 50\,$\Omega$.
\begin{figure}[h]
    \noindent\centering{
    \includegraphics[width=75mm]{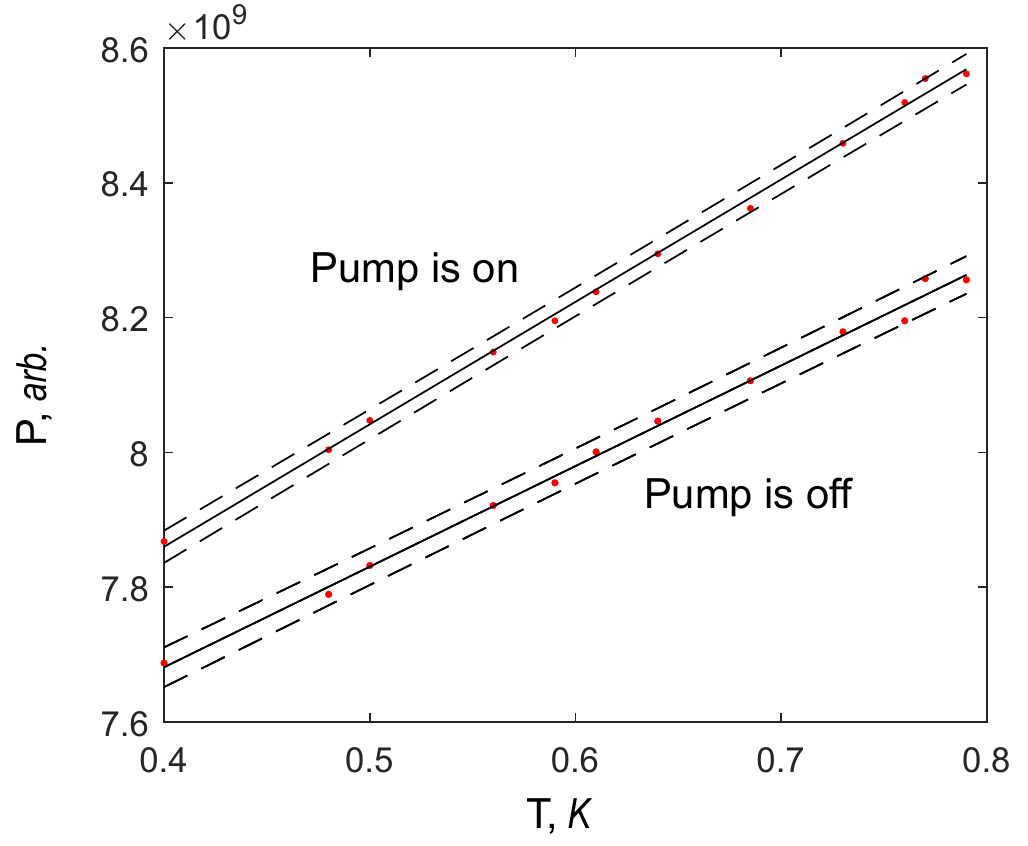}
    }
    \caption{
    Noise power at different temperatures: experiment (red dots), linear fit (solid black line), and variance (dashed black lines).
    The linear fit provides the system gain when TWPA is on $G_{ON}$ and gain when TWPA is off $G_{OFF}$, the difference of which gives the TWPA's gain of $G_T = 0.853$\,dB.
    }
    \label{calibration}
\end{figure}

We calculate the system noise temperature by dividing the noise temperature in the case when TWPA is turned off $T_{N_{OFF}}$ by the signal-to-noise-ratio improvement using the following formula \cite{Roy2015}
\begin{equation}
    T_{sys} = T_{N_{OFF}}\,(\delta_{SNR}^{-1}-G_{T}^{-1}),
\end{equation}
where $\delta_{SNR}$ is the signal-to-noise ratio improvement and $G_{T}$ is TWPA's gain.

In order to calibrate the $I$ and $Q$ variances for the entanglement verification, we measure the voltage variance when the TWPA is off and extract it from the variance measured in the case TWPA generates photons.
Covariances are calculated by normalization per one photon.
For the theoretical analysis, the measured voltage quadratures are calibrated and scaled according to the results of prior measurements of the system gain and noise temperature.
Using a linear fit of the noise power as function of temperature, we observed that in the case of unpumped TWPA $G_{OFF} = 91.74 \pm 0.20$\,dB (see pump-off case in Fig.\,\ref{calibration}) and $T_{sys} = 4.75$\,K.
We observe that the gain varies with time and the variations are in the range of $\pm 0.2$\,dB.

\section{Loss model: Using the TWPA as an amplifier}

To formulate a distributed model for a lossy waveguide with gain, a general mathematical description has been introduced in Ref. \onlinecite{Haus2000}. According to this model, each section of the waveguide contains loss per length $\mathcal{L}$ and gain per length $\mathcal{G}$. As the signal propagates along the device, noise will be added, it will get amplified, and will simultaneously decay into the loss channel. Considering the parametric dynamics of a general TWPA, this simple model is addressed further in Ref \onlinecite{Houde2019}, which will be also employed in the following.

Consider perfect phase-matching but include distributed loss with identical frequency independent coefficient $\kappa$ along the TWPA. The Heisenberg-Langevin equation for the signal mode $\hat{a}_S$ and idler mode $\hat{a}_I$ inside the TWPA with the length $L$ is then given by [\onlinecite{Houde2019}]
\begin{eqnarray}
(\partial_t + v \partial_x)\hat{a}_S(x) = \chi \hat{a}_I^{\dagger}(x) -\frac{\kappa}{2}\hat{a}_S(x) + \sqrt{\kappa}\hat{\xi}_S(x),\nonumber
\\
(\partial_t + v \partial_x)\hat{a}_I(x)^{\dagger} = \chi \hat{a}_S(x) -\frac{\kappa}{2}\hat{a}_I^{\dagger}(x) + \sqrt{\kappa}\hat{\xi}_I^{\dagger}(x),
\end{eqnarray}
where $v$ is the group velocity, $\chi$ is the parametric interaction strength, and $\hat{\xi}_i$ are noise mode sources. After moving to the frequency domain, one can find a scattering matrix $\mathbf{S}(x)$ linking the modes at the output of the TWPA, $x=L$, to those at $x=0$ [\onlinecite{Houde2019}]:
\begin{equation}\label{SI_eq3}
(\hat{a}_s[L,\omega],\hat{a}^{\dagger}_I[L,\omega])^T=\mathbf{S}(L)(\hat{a}_s[0,\omega],\hat{a}_I^{\dagger}[0,\omega])^T+\frac{\sqrt{\kappa}}{v}\int_0^L dx'\, \mathbf{S}(L-x')(\hat{\xi}_s,\hat{\xi}_I^{\dagger})^T,
\end{equation}
where $(x,y)^T$ denotes a vertical vector with elements $x$ and $y$. Here elements of the scattering matrix are given by
\begin{eqnarray}
S_{11}(x)=S_{22}(x)=e^{-\frac{\kappa \chi}{2v}+i \omega }\cosh{\frac{\chi x}{v}},\nonumber
\\
S_{12}(x)=S_{21}(x)=e^{-\frac{\kappa \chi}{2v}+i \omega }\sinh{\frac{\chi x}{v}}.
\end{eqnarray}
This clearly represents simultaneous gain and loss along the TWPA. In the ideal case that the TPWA is lossless ($\kappa=0$), one recovers the well-known relation (excluding the global phase):
\begin{equation}
\hat{a}_S[L,\omega]=\cosh(r)\hat{a}_S[0,\omega]+\sinh(r)\hat{a}^{\dagger}_I[0,\omega],  
\end{equation}
with $r=\chi L/v$. However, when $\kappa$ is nonzero the noise sources will be added to the propagating modes and couple due to the parametric interaction, i.e. noise added to the signal contributes to the idler mode and vice versa. For the sake of simplicity, one can adapt a lumped-element picture of Eq. \ref{SI_eq3} to link the input and output signals of the TWPA. Assuming symmetric loss of signal and idler, one gets
\begin{eqnarray}\label{SI_eq6}
\hat{a}_{out,S}&=& \sqrt{\eta G_S}\,\hat{a}_{in,S}+\sqrt{\eta G_I}\,\hat{a}_{in,I}^{\dagger} + \sqrt{G_S(1-\eta)}\,\hat{\xi}_{S} + \sqrt{G_I(1-\eta)}\,\hat{\xi}^{\dagger}_{I},
\end{eqnarray}
where $G_{S}$ and $G_{I}$ describe the gain of the TWPA at signal's and idler's frequencies and $\eta$ corresponds to its loss.

The internal losses of TWPA are modeled by the shunt capacitance loss tangent. 
From the process characterization of the atomic layer aluminum oxide deposition we estimate  $\tan{(\delta)}=0.0025$.
The total loss in the waveguide is $e^{-\tan{(\delta)} \theta}$, where $\theta$ is electrical length of TWPA in radians. 
The estimated electrical length varies in range $\theta=60\pm15$ rads at $\Phi_{DC}=0.38\Phi_0$ that gives $\eta=-0.65\pm0.20$\,dB of internal losses that we use in the calibration.
Such a loss value coincides with the one measured in the experiment with the same sample.
In that experiment, two microwave switches were used to create two paths: one through the TWPA and one through the resonator placed in the same TWPA's sample holder.
{\color{black} Over the whole $4-8$\,GHz bandwidth, the average internal loss $\eta\approx-1.18$\,dB.}

\section{on-off calibration}
In this section we discuss the on-off calibration technique for obtaining the covariance matrix of the output modes of the TWPA based on the covariance matrix of the measured signals at the end of the system [\onlinecite{Eichler2011, Flurin2015}]. 
We define $K=\{\hat{I}_S,\hat{Q}_S,\hat{I}_I,\hat{Q}_I\}$, where the quadratures are given by $\hat{I}_i=(\hat{a}_i+\hat{a}^{\dagger}_i)/2$ and $\hat{Q}_i=-i(\hat{a}_i-\hat{a}^{\dagger}_i)/2$, with the index $i=S,I$. 
We note that by considering these definitions of the quadratures we relate to the real and the imaginary parts of the field's amplitudes, which we actually measure in the experiment.
Assuming zero-mean modes, our aim is to calibrate the covariance matrix with elements $V_{ij}=\langle \hat{K}_i\hat{K}_j+\hat{K}_j\hat{K}_i\rangle /2$ at the output of the TWPA since the covariance matrix fully characterizes the Gaussian states.
We prove the Gaussian nature of the output signals by measuring the skewness $\langle \delta V_{IQ}^3 \rangle/\sigma^3$ and kurtosis $\langle \delta V_{IQ}^4 \rangle/\sigma^4$ of the signal $V_{IQ}$, where $\sigma$ is the standard deviation and $\delta V_{IQ}=V_{IQ}-\langle V_{IQ} \rangle$.
The observed skewness and kurtosis are $0.001\pm0.005$ and $2.998\pm0.003$ respectively, which proves the states are, indeed, Gaussian states.
\begin{figure}[h]
    \noindent\centering{
    \includegraphics[width=125mm]{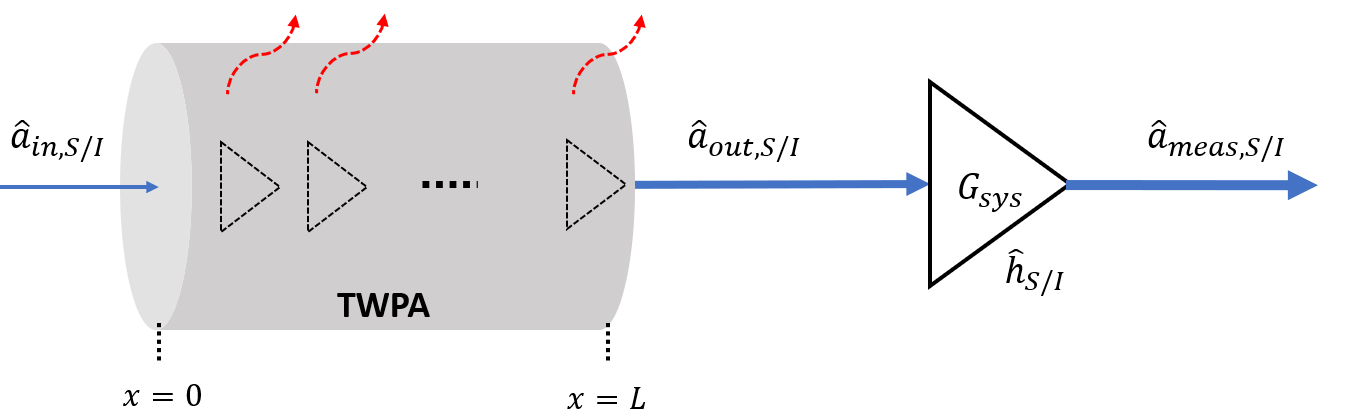}
    }
    \caption{
    Schematic description of the system chain. The TWPA is modeled by distributed gain and loss, which results in an effective power-dependent gain denoted by $G_T(P_p)$, with $P_p$ denoting the power of the pump. The output modes of the TWPA include the two-mode squeezed vacuum plus added noise sources due to loss of the device. Before reaching the measurement stage, signals get amplified by the HEMT and room temperature amplifiers (collectively described by the system gain $G_{sys}$ and uncorrelated noise sources $\hat{h}_{S/I}$). Thus the combined gain of the system chain is described by $G_{ON}(P_p)=G_T(P_p)G_{sys}.$
    }
    \label{amplififer_chain}
\end{figure}

Suppose that the competition between distributed loss and gain along the TWPA results in an effective power-dependent gain $G_T(P_p)$, such that by referring to Eq. \ref{SI_eq6} one finds $G_T(P_p):=\eta G_S$. Thus, when the TWPA is pumped the gain can compensate the loss and one has $G_T(P_p)>1$. However, when the pump is off the loss can not be compensated and we get the attenuation $G_T(P_p=0)<1$. Assuming that the input of the TWPA are vacuum fields $\hat{v}_{S/I}$, the measured fields in presence of pump ($P_p\neq 0$) read
\begin{eqnarray}
\hat{a}_{meas,S}^{(ON)}=\sqrt{G_{sys}}\big(\sqrt{\eta G_S}\,\hat{v}_S+\sqrt{\eta G_I}\,\hat{v}_I^{\dagger} + \sqrt{G_S(1-\eta)}\,\hat{\xi}_{S} + \sqrt{G_I(1-\eta)}\,\hat{\xi}^{\dagger}_{I}\big)+\sqrt{G_{sys}-1}\hat{h}_S^{\dagger},
\\
\hat{a}_{meas,I}^{(ON)}=\sqrt{G_{sys}}\big(\sqrt{\eta G_I}\,\hat{v}_I+\sqrt{\eta G_S}\,\hat{v}_S^{\dagger} + \sqrt{G_I(1-\eta)}\,\hat{\xi}_{I} + \sqrt{G_S(1-\eta)}\,\hat{\xi}^{\dagger}_{S}\big)+\sqrt{G_{sys}-1}\hat{h}_I^{\dagger}.
\end{eqnarray}
If the TWPA is not pumped, then the overall gain of the TWPA is described by the attenuation $G_T(P_p=0)=\eta$. Considering the noise sources $\hat{\zeta}_{S/I}$ added due to this attenuation, one obtains 
\begin{eqnarray}
\hat{a}_{meas,S}^{(OFF)}=\sqrt{G_{sys}\,\eta}\,\hat{v}_S+\sqrt{G_{sys}\,(1-\eta)}\,\hat{\zeta}_S+\sqrt{G_{sys}-1}\hat{h}_S^{\dagger},\label{SI_eq9}
\\
\hat{a}_{meas,I}^{(OFF)}=\sqrt{G_{sys}\,\eta}\,\hat{v}_I+\sqrt{G_{sys}\,(1-\eta)}\,\hat{\zeta}_I+\sqrt{G_{sys}-1}\hat{h}_I^{\dagger}.\label{SI_eq10}
\end{eqnarray}
In further analysis of variances and covariances, we omit all coefficients in normalization $\mathcal{N}$ except for the gain.
Thus, subtracting the covariance matrix of the measured quadratures when the pump is on from the covariance matrix of the measured quadratures when the pump is off yields 
\begin{equation}
\mathbf{V}_{meas}^{(ON)}-\mathbf{V}_{meas}^{(OFF)} = G_{sys}\mathbf{V}_{out} - \frac{G_{sys}\eta }{4}\mathbf{I}_4 ~  - G_{sys}(1-\eta)\mathbf{V}_{noise},
\end{equation}
where $\mathbf{V}_{out}$ denotes the covariance matrix of the output modes of TWPA, the identity matrix is related to the covariance matrix of the vacuum inputs, and $\mathbf{V}_{noise}$ denotes the added noise of the unpumped TWPA due to its loss. 
Accordingly, one finds
\begin{equation}
\mathbf{V}_{out}=\frac{\mathbf{V}_{meas}^{(ON)}-\mathbf{V}_{meas}^{(OFF)}}{G_{sys}}+\frac{\eta}{4}\mathbf{I}_4+(1-\eta)\mathbf{V}_{noise}.
\end{equation}

Recalling that $\eta=G_T(P_p=0)$, $G_{OFF}=\eta G_{sys}$ and assuming perfect thermalization for the TWPA's noise sources in Eqs.(\ref{SI_eq9}-\ref{SI_eq10}), we obtain
\begin{equation}
\mathbf{V}_{out} =  \frac{\eta\big(\mathbf{V}_{meas}^{(ON)}-\mathbf{V}_{meas}^{(OFF)}\big)}{G_{OFF}} + \frac{1}{4}\mathbf{I}_4.
\end{equation}
By returning back all normalization coefficients, in addition to multiplying $\mathbf{V}_{out}$ by a factor $4$ -- to rescale the vacuum fluctuations to $1$ (instead of $1/4$) -- we get the expressions given in Eq.~(3) of the main text.

\section{Error analysis}
The two main sources of error in the estimation of entanglement strength and squeezing are inaccuracies in the gain of the system and in the internal losses of the TWPA.
The system gain is $G_{OFF} = 91.74 \pm 0.20$\,dB and the loss coefficient is $\eta = -0.65 \pm 0.20$\,dB.
This gives an uncertainty of $\pm0.4$\,dB for the normalization coefficient $\mathcal{N}$.
Taking the highest squeezing point we obtain $\nu = 0.55 \pm 0.10$, which is given by the error in normalization coefficients.

\section{Rate of Entanglement generation and purity}

To quantify the entanglement generation rate of TWPA we introduce $R_E = 2 \langle \mathcal{I}_1^2 \rangle\,E_F (\Delta\omega+\delta\omega)$.
The photon flux intensity $\langle \mathcal{I}_1^2 \rangle$ is extracted directly from the experiment and $(\Delta\omega+\delta\omega)$ is set manually.
The entropy of formation $E_F$ according to Refs.\,\onlinecite{Laurat2005, Flurin2012} can be calculated from the symplectic eigenvalue using the following relation
\begin{equation}
    E_F = c_+\log{c_+}-c_-\log{c_-},
\end{equation}
where $c_{\pm}=(\nu_{min}^{-1/2} \pm \nu_{min}^{1/2})/4$.

The purity of the quantum state is calculated from the covariance matrix $\mathbf{V}_{out}$: $\mu=1/\sqrt{|\det(\mathbf{V}_{out})|}$, were $\det$ is the matrix determinant. 

{\color{black}
\section{Single-mode squeezing}

Besides two-mode entanglement that we characterize with the covariance matrix depicted in Fig.\,\ref{single_squeezing}a, we demonstrate the generation of single-mode squeezed states. 
For this experiment, we consider a spanned single mode with 0.5M\,Hz bandwidth centered at $\omega_p/2$. 
The parametric amplifier operates in the degenerate mode at this point.
These amplitudes are shown in Fig.\,\ref{single_squeezing}b as function of the pump power. 
The maximum squeezing of $2.4 \pm 0.7$\,dB is achieved at $-68$\,dBm pump power.

\begin{figure}[h]
    \noindent\centering{
    \includegraphics[width=135mm]{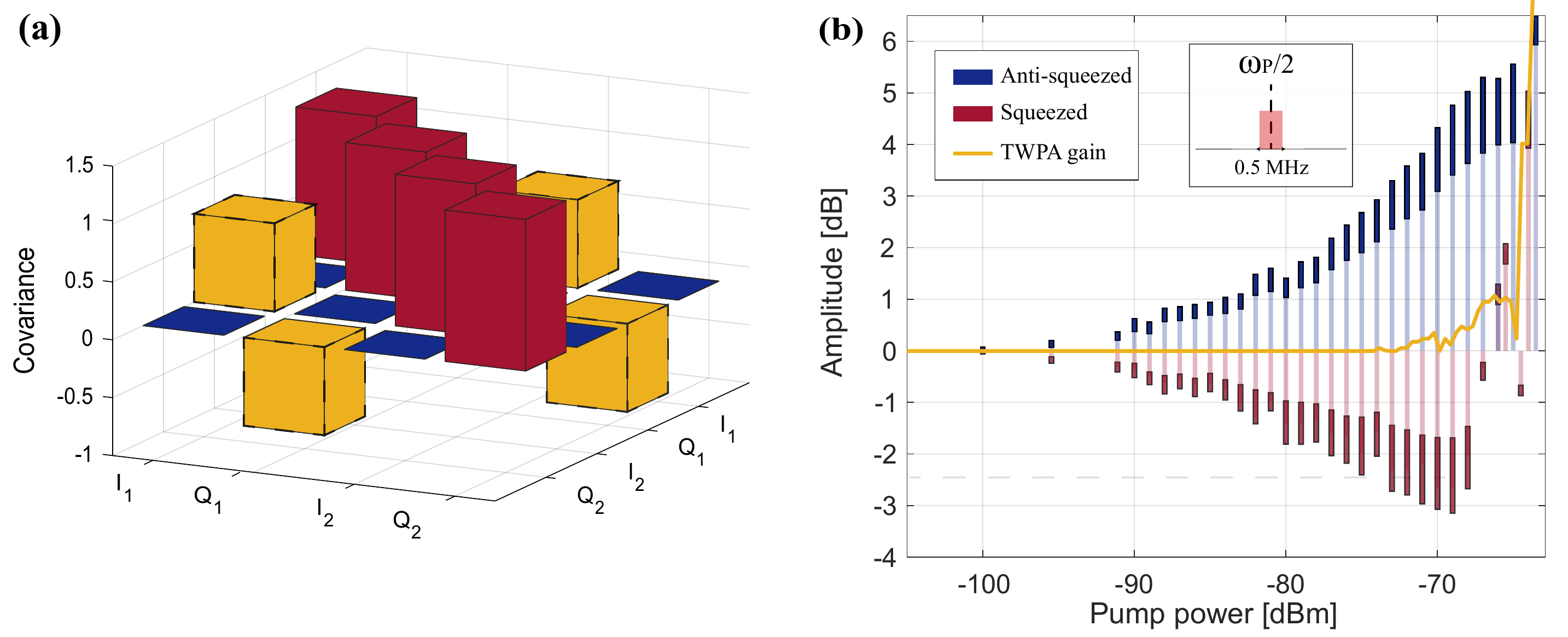}
    }
    \caption{
    (a) The covariance matrix for the two-mode entangled state.
    (b) The single-mode squeezing for different pump powers. 
    Transparent blue bars corresponds to the measured amplified quadrature, while transparent red bars corresponds to the measured squeezed quadrature. 
    The solid red and blue bars correspond to the errors in the amplitude estimation. The yellow curve shows the gain of the TWPA.
    }
    \label{single_squeezing}
\end{figure}
}
\bibliography{apssamp}

\providecommand{\noopsort}[1]{}\providecommand{\singleletter}[1]{#1}%
\begin{thebibliography}{77}%
\makeatletter
\providecommand \@ifxundefined [1]{%
 \@ifx{#1\undefined}
}%
\providecommand \@ifnum [1]{%
 \ifnum #1\expandafter \@firstoftwo
 \else \expandafter \@secondoftwo
 \fi
}%
\providecommand \@ifx [1]{%
 \ifx #1\expandafter \@firstoftwo
 \else \expandafter \@secondoftwo
 \fi
}%
\providecommand \natexlab [1]{#1}%
\providecommand \enquote  [1]{``#1''}%
\providecommand \bibnamefont  [1]{#1}%
\providecommand \bibfnamefont [1]{#1}%
\providecommand \citenamefont [1]{#1}%
\providecommand \href@noop [0]{\@secondoftwo}%
\providecommand \href [0]{\begingroup \@sanitize@url \@href}%
\providecommand \@href[1]{\@@startlink{#1}\@@href}%
\providecommand \@@href[1]{\endgroup#1\@@endlink}%
\providecommand \@sanitize@url [0]{\catcode `\\12\catcode `\$12\catcode
  `\&12\catcode `\#12\catcode `\^12\catcode `\_12\catcode `\%12\relax}%
\providecommand \@@startlink[1]{}%
\providecommand \@@endlink[0]{}%
\providecommand \url  [0]{\begingroup\@sanitize@url \@url }%
\providecommand \@url [1]{\endgroup\@href {#1}{\urlprefix }}%
\providecommand \urlprefix  [0]{URL }%
\providecommand \Eprint [0]{\href }%
\providecommand \doibase [0]{http://dx.doi.org/}%
\providecommand \selectlanguage [0]{\@gobble}%
\providecommand \bibinfo  [0]{\@secondoftwo}%
\providecommand \bibfield  [0]{\@secondoftwo}%
\providecommand \translation [1]{[#1]}%
\providecommand \BibitemOpen [0]{}%
\providecommand \bibitemStop [0]{}%
\providecommand \bibitemNoStop [0]{.\EOS\space}%
\providecommand \EOS [0]{\spacefactor3000\relax}%
\providecommand \BibitemShut  [1]{\csname bibitem#1\endcsname}%
\let\auto@bib@innerbib\@empty
\bibitem [{\citenamefont {Brecht}\ \emph {et~al.}(2016)\citenamefont {Brecht},
  \citenamefont {Pfaff}, \citenamefont {Wang}, \citenamefont {Chu},
  \citenamefont {Frunzio}, \citenamefont {Devoret},\ and\ \citenamefont
  {Schoelkopf}}]{Brecht2016}%
  \BibitemOpen
  \bibfield  {author} {\bibinfo {author} {\bibfnamefont {T.}~\bibnamefont
  {Brecht}}, \bibinfo {author} {\bibfnamefont {W.}~\bibnamefont {Pfaff}},
  \bibinfo {author} {\bibfnamefont {C.}~\bibnamefont {Wang}}, \bibinfo {author}
  {\bibfnamefont {Y.}~\bibnamefont {Chu}}, \bibinfo {author} {\bibfnamefont
  {L.}~\bibnamefont {Frunzio}}, \bibinfo {author} {\bibfnamefont {M.~H.}\
  \bibnamefont {Devoret}}, \ and\ \bibinfo {author} {\bibfnamefont {R.~J.}\
  \bibnamefont {Schoelkopf}},\ }\href {\doibase 10.1038/npjqi.2016.2}
  {\bibfield  {journal} {\bibinfo  {journal} {npj Quantum Information}\
  }\textbf {\bibinfo {volume} {2}},\ \bibinfo {pages} {16002} (\bibinfo {year}
  {2016})}\BibitemShut {NoStop}%
\bibitem [{\citenamefont {Magnard}\ \emph {et~al.}(2020)\citenamefont
  {Magnard}, \citenamefont {Storz}, \citenamefont {Kurpiers}, \citenamefont
  {Sch\"ar}, \citenamefont {Marxer}, \citenamefont {L\"utolf}, \citenamefont
  {Walter}, \citenamefont {Besse}, \citenamefont {Gabureac}, \citenamefont
  {Reuer}, \citenamefont {Akin}, \citenamefont {Royer}, \citenamefont {Blais},\
  and\ \citenamefont {Wallraff}}]{Magnard2020}%
  \BibitemOpen
  \bibfield  {author} {\bibinfo {author} {\bibfnamefont {P.}~\bibnamefont
  {Magnard}}, \bibinfo {author} {\bibfnamefont {S.}~\bibnamefont {Storz}},
  \bibinfo {author} {\bibfnamefont {P.}~\bibnamefont {Kurpiers}}, \bibinfo
  {author} {\bibfnamefont {J.}~\bibnamefont {Sch\"ar}}, \bibinfo {author}
  {\bibfnamefont {F.}~\bibnamefont {Marxer}}, \bibinfo {author} {\bibfnamefont
  {J.}~\bibnamefont {L\"utolf}}, \bibinfo {author} {\bibfnamefont
  {T.}~\bibnamefont {Walter}}, \bibinfo {author} {\bibfnamefont {J.-C.}\
  \bibnamefont {Besse}}, \bibinfo {author} {\bibfnamefont {M.}~\bibnamefont
  {Gabureac}}, \bibinfo {author} {\bibfnamefont {K.}~\bibnamefont {Reuer}},
  \bibinfo {author} {\bibfnamefont {A.}~\bibnamefont {Akin}}, \bibinfo {author}
  {\bibfnamefont {B.}~\bibnamefont {Royer}}, \bibinfo {author} {\bibfnamefont
  {A.}~\bibnamefont {Blais}}, \ and\ \bibinfo {author} {\bibfnamefont
  {A.}~\bibnamefont {Wallraff}},\ }\href {\doibase
  10.1103/PhysRevLett.125.260502} {\bibfield  {journal} {\bibinfo  {journal}
  {Physical Review Letters}\ }\textbf {\bibinfo {volume} {125}},\ \bibinfo
  {pages} {260502} (\bibinfo {year} {2020})}\BibitemShut {NoStop}%
\bibitem [{\citenamefont {Burkhart}\ \emph {et~al.}(2020)\citenamefont
  {Burkhart}, \citenamefont {Teoh}, \citenamefont {Zhang}, \citenamefont
  {Axline}, \citenamefont {Frunzio}, \citenamefont {Devoret}, \citenamefont
  {Jiang}, \citenamefont {Girvin},\ and\ \citenamefont
  {Schoelkopf}}]{Burkhart2020}%
  \BibitemOpen
  \bibfield  {author} {\bibinfo {author} {\bibfnamefont {L.~D.}\ \bibnamefont
  {Burkhart}}, \bibinfo {author} {\bibfnamefont {J.}~\bibnamefont {Teoh}},
  \bibinfo {author} {\bibfnamefont {Y.}~\bibnamefont {Zhang}}, \bibinfo
  {author} {\bibfnamefont {C.~J.}\ \bibnamefont {Axline}}, \bibinfo {author}
  {\bibfnamefont {L.}~\bibnamefont {Frunzio}}, \bibinfo {author} {\bibfnamefont
  {M.}~\bibnamefont {Devoret}}, \bibinfo {author} {\bibfnamefont
  {L.}~\bibnamefont {Jiang}}, \bibinfo {author} {\bibfnamefont
  {S.}~\bibnamefont {Girvin}}, \ and\ \bibinfo {author} {\bibfnamefont
  {R.}~\bibnamefont {Schoelkopf}},\ }\href@noop {} {\enquote {\bibinfo {title}
  {Error-detected state transfer and entanglement in a superconducting quantum
  network},}\ } (\bibinfo {year} {2020}),\ \Eprint {http://arxiv.org/abs/arXiv
  2004.06168} {arXiv:arXiv 2004.06168 [quant-ph]} \BibitemShut {NoStop}%
\bibitem [{\citenamefont {Lloyd}(2008)}]{Lloyd2008}%
  \BibitemOpen
  \bibfield  {author} {\bibinfo {author} {\bibfnamefont {S.}~\bibnamefont
  {Lloyd}},\ }\href {\doibase 10.1126/science.1160627} {\bibfield  {journal}
  {\bibinfo  {journal} {Science}\ }\textbf {\bibinfo {volume} {321}},\ \bibinfo
  {pages} {1463} (\bibinfo {year} {2008})}\BibitemShut {NoStop}%
\bibitem [{\citenamefont {Barzanjeh}\ \emph {et~al.}(2015)\citenamefont
  {Barzanjeh}, \citenamefont {Guha}, \citenamefont {Weedbrook}, \citenamefont
  {Vitali}, \citenamefont {Shapiro},\ and\ \citenamefont
  {Pirandola}}]{Barzanjeh2015}%
  \BibitemOpen
  \bibfield  {author} {\bibinfo {author} {\bibfnamefont {S.}~\bibnamefont
  {Barzanjeh}}, \bibinfo {author} {\bibfnamefont {S.}~\bibnamefont {Guha}},
  \bibinfo {author} {\bibfnamefont {C.}~\bibnamefont {Weedbrook}}, \bibinfo
  {author} {\bibfnamefont {D.}~\bibnamefont {Vitali}}, \bibinfo {author}
  {\bibfnamefont {J.~H.}\ \bibnamefont {Shapiro}}, \ and\ \bibinfo {author}
  {\bibfnamefont {S.}~\bibnamefont {Pirandola}},\ }\href {\doibase
  10.1103/PhysRevLett.114.080503} {\bibfield  {journal} {\bibinfo  {journal}
  {Physical Review Letters}\ }\textbf {\bibinfo {volume} {114}},\ \bibinfo
  {pages} {080503} (\bibinfo {year} {2015})}\BibitemShut {NoStop}%
\bibitem [{\citenamefont {Pogorzalek}\ \emph {et~al.}(2019)\citenamefont
  {Pogorzalek}, \citenamefont {Fedorov}, \citenamefont {Xu}, \citenamefont
  {Parra-Rodriguez}, \citenamefont {Sanz}, \citenamefont {Fischer},
  \citenamefont {Xie}, \citenamefont {Inomata}, \citenamefont {Nakamura},
  \citenamefont {Solano}, \citenamefont {Marx}, \citenamefont {Deppe},\ and\
  \citenamefont {Gross}}]{Pogorzalek2019}%
  \BibitemOpen
  \bibfield  {author} {\bibinfo {author} {\bibfnamefont {S.}~\bibnamefont
  {Pogorzalek}}, \bibinfo {author} {\bibfnamefont {K.~G.}\ \bibnamefont
  {Fedorov}}, \bibinfo {author} {\bibfnamefont {M.}~\bibnamefont {Xu}},
  \bibinfo {author} {\bibfnamefont {A.}~\bibnamefont {Parra-Rodriguez}},
  \bibinfo {author} {\bibfnamefont {M.}~\bibnamefont {Sanz}}, \bibinfo {author}
  {\bibfnamefont {M.}~\bibnamefont {Fischer}}, \bibinfo {author} {\bibfnamefont
  {E.}~\bibnamefont {Xie}}, \bibinfo {author} {\bibfnamefont {K.}~\bibnamefont
  {Inomata}}, \bibinfo {author} {\bibfnamefont {Y.}~\bibnamefont {Nakamura}},
  \bibinfo {author} {\bibfnamefont {E.}~\bibnamefont {Solano}}, \bibinfo
  {author} {\bibfnamefont {A.}~\bibnamefont {Marx}}, \bibinfo {author}
  {\bibfnamefont {F.}~\bibnamefont {Deppe}}, \ and\ \bibinfo {author}
  {\bibfnamefont {R.}~\bibnamefont {Gross}},\ }\href {\doibase
  10.1038/s41467-019-10727-7} {\bibfield  {journal} {\bibinfo  {journal}
  {Nature Communications}\ }\textbf {\bibinfo {volume} {10}},\ \bibinfo {pages}
  {2604} (\bibinfo {year} {2019})}\BibitemShut {NoStop}%
\bibitem [{\citenamefont {Candia}\ \emph {et~al.}(2020)\citenamefont {Candia},
  \citenamefont {Yiğitler}, \citenamefont {Paraoanu},\ and\ \citenamefont
  {Jäntti}}]{DiCandia2020}%
  \BibitemOpen
  \bibfield  {author} {\bibinfo {author} {\bibfnamefont {R.~D.}\ \bibnamefont
  {Candia}}, \bibinfo {author} {\bibfnamefont {H.}~\bibnamefont {Yiğitler}},
  \bibinfo {author} {\bibfnamefont {G.~S.}\ \bibnamefont {Paraoanu}}, \ and\
  \bibinfo {author} {\bibfnamefont {R.}~\bibnamefont {Jäntti}},\ }\href@noop
  {} {\enquote {\bibinfo {title} {Two-way covert microwave quantum
  communication},}\ } (\bibinfo {year} {2020}),\ \Eprint
  {http://arxiv.org/abs/arXiv 2004.07192} {arXiv:arXiv 2004.07192 [quant-ph]}
  \BibitemShut {NoStop}%
\bibitem [{\citenamefont {Fedorov}\ \emph {et~al.}(2021)\citenamefont
  {Fedorov}, \citenamefont {Renger}, \citenamefont {Pogorzalek}, \citenamefont
  {Candia}, \citenamefont {Chen}, \citenamefont {Nojiri}, \citenamefont
  {Inomata}, \citenamefont {Nakamura}, \citenamefont {Partanen}, \citenamefont
  {Marx}, \citenamefont {Gross},\ and\ \citenamefont {Deppe}}]{Fedorov2021}%
  \BibitemOpen
  \bibfield  {author} {\bibinfo {author} {\bibfnamefont {K.~G.}\ \bibnamefont
  {Fedorov}}, \bibinfo {author} {\bibfnamefont {M.}~\bibnamefont {Renger}},
  \bibinfo {author} {\bibfnamefont {S.}~\bibnamefont {Pogorzalek}}, \bibinfo
  {author} {\bibfnamefont {R.~D.}\ \bibnamefont {Candia}}, \bibinfo {author}
  {\bibfnamefont {Q.}~\bibnamefont {Chen}}, \bibinfo {author} {\bibfnamefont
  {Y.}~\bibnamefont {Nojiri}}, \bibinfo {author} {\bibfnamefont
  {K.}~\bibnamefont {Inomata}}, \bibinfo {author} {\bibfnamefont
  {Y.}~\bibnamefont {Nakamura}}, \bibinfo {author} {\bibfnamefont
  {M.}~\bibnamefont {Partanen}}, \bibinfo {author} {\bibfnamefont
  {A.}~\bibnamefont {Marx}}, \bibinfo {author} {\bibfnamefont {R.}~\bibnamefont
  {Gross}}, \ and\ \bibinfo {author} {\bibfnamefont {F.}~\bibnamefont
  {Deppe}},\ }\href@noop {} {\enquote {\bibinfo {title} {Experimental quantum
  teleportation of propagating microwaves},}\ } (\bibinfo {year} {2021}),\
  \Eprint {http://arxiv.org/abs/arXiv 2103.04155} {arXiv:arXiv 2103.04155
  [quant-ph]} \BibitemShut {NoStop}%
\bibitem [{\citenamefont {Yurke}\ \emph {et~al.}(1988)\citenamefont {Yurke},
  \citenamefont {Kaminsky}, \citenamefont {Miller}, \citenamefont {Whittaker},
  \citenamefont {Smith}, \citenamefont {Silver},\ and\ \citenamefont
  {Simon}}]{Yurke1988}%
  \BibitemOpen
  \bibfield  {author} {\bibinfo {author} {\bibfnamefont {B.}~\bibnamefont
  {Yurke}}, \bibinfo {author} {\bibfnamefont {P.~G.}\ \bibnamefont {Kaminsky}},
  \bibinfo {author} {\bibfnamefont {R.~E.}\ \bibnamefont {Miller}}, \bibinfo
  {author} {\bibfnamefont {E.~A.}\ \bibnamefont {Whittaker}}, \bibinfo {author}
  {\bibfnamefont {A.~D.}\ \bibnamefont {Smith}}, \bibinfo {author}
  {\bibfnamefont {A.~H.}\ \bibnamefont {Silver}}, \ and\ \bibinfo {author}
  {\bibfnamefont {R.~W.}\ \bibnamefont {Simon}},\ }\href {\doibase
  10.1103/physrevlett.60.764} {\bibfield  {journal} {\bibinfo  {journal}
  {Physical Review Letters}\ }\textbf {\bibinfo {volume} {60}},\ \bibinfo
  {pages} {764} (\bibinfo {year} {1988})}\BibitemShut {NoStop}%
\bibitem [{\citenamefont {Yamamoto}\ \emph {et~al.}(2008)\citenamefont
  {Yamamoto}, \citenamefont {Inomata}, \citenamefont {Watanabe}, \citenamefont
  {Matsuba}, \citenamefont {Miyazaki}, \citenamefont {Oliver}, \citenamefont
  {Nakamura},\ and\ \citenamefont {Tsai}}]{Yamamoto2008}%
  \BibitemOpen
  \bibfield  {author} {\bibinfo {author} {\bibfnamefont {T.}~\bibnamefont
  {Yamamoto}}, \bibinfo {author} {\bibfnamefont {K.}~\bibnamefont {Inomata}},
  \bibinfo {author} {\bibfnamefont {M.}~\bibnamefont {Watanabe}}, \bibinfo
  {author} {\bibfnamefont {K.}~\bibnamefont {Matsuba}}, \bibinfo {author}
  {\bibfnamefont {T.}~\bibnamefont {Miyazaki}}, \bibinfo {author}
  {\bibfnamefont {W.~D.}\ \bibnamefont {Oliver}}, \bibinfo {author}
  {\bibfnamefont {Y.}~\bibnamefont {Nakamura}}, \ and\ \bibinfo {author}
  {\bibfnamefont {J.~S.}\ \bibnamefont {Tsai}},\ }\href {\doibase
  10.1063/1.2964182} {\bibfield  {journal} {\bibinfo  {journal} {Applied
  Physical Letters}\ }\textbf {\bibinfo {volume} {93}},\ \bibinfo {pages}
  {042510} (\bibinfo {year} {2008})}\BibitemShut {NoStop}%
\bibitem [{\citenamefont {Hatridge}\ \emph {et~al.}(2011)\citenamefont
  {Hatridge}, \citenamefont {Vijay}, \citenamefont {Slichter}, \citenamefont
  {Clarke},\ and\ \citenamefont {Siddiqi}}]{Hatridge2011}%
  \BibitemOpen
  \bibfield  {author} {\bibinfo {author} {\bibfnamefont {M.}~\bibnamefont
  {Hatridge}}, \bibinfo {author} {\bibfnamefont {R.}~\bibnamefont {Vijay}},
  \bibinfo {author} {\bibfnamefont {D.~H.}\ \bibnamefont {Slichter}}, \bibinfo
  {author} {\bibfnamefont {J.}~\bibnamefont {Clarke}}, \ and\ \bibinfo {author}
  {\bibfnamefont {I.}~\bibnamefont {Siddiqi}},\ }\href {\doibase
  10.1103/PhysRevB.83.134501} {\bibfield  {journal} {\bibinfo  {journal}
  {Physical Review B}\ }\textbf {\bibinfo {volume} {83}},\ \bibinfo {pages}
  {134501} (\bibinfo {year} {2011})}\BibitemShut {NoStop}%
\bibitem [{\citenamefont {L{\"{a}}hteenm{\"{a}}ki}\ \emph
  {et~al.}(2012)\citenamefont {L{\"{a}}hteenm{\"{a}}ki}, \citenamefont
  {Vesterinen}, \citenamefont {Hassel}, \citenamefont {Sepp{\"{a}}},\ and\
  \citenamefont {Hakonen}}]{Lahteenmaki2012}%
  \BibitemOpen
  \bibfield  {author} {\bibinfo {author} {\bibfnamefont {P.}~\bibnamefont
  {L{\"{a}}hteenm{\"{a}}ki}}, \bibinfo {author} {\bibfnamefont
  {V.}~\bibnamefont {Vesterinen}}, \bibinfo {author} {\bibfnamefont
  {J.}~\bibnamefont {Hassel}}, \bibinfo {author} {\bibfnamefont
  {H.}~\bibnamefont {Sepp{\"{a}}}}, \ and\ \bibinfo {author} {\bibfnamefont
  {P.}~\bibnamefont {Hakonen}},\ }\href {\doibase 10.1038/srep00276} {\bibfield
   {journal} {\bibinfo  {journal} {Sci. Rep.}\ }\textbf {\bibinfo {volume}
  {2}},\ \bibinfo {pages} {276} (\bibinfo {year} {2012})}\BibitemShut {NoStop}%
\bibitem [{\citenamefont {Mutus}\ \emph {et~al.}(2013)\citenamefont {Mutus},
  \citenamefont {White}, \citenamefont {Jeffrey}, \citenamefont {Sank},
  \citenamefont {Barends}, \citenamefont {Bochmann}, \citenamefont {Chen},
  \citenamefont {Chen}, \citenamefont {Chiaro}, \citenamefont {Dunsworth},
  \citenamefont {Kelly}, \citenamefont {Megrant}, \citenamefont {Neill},
  \citenamefont {O'Malley}, \citenamefont {Roushan}, \citenamefont
  {Vainsencher}, \citenamefont {Wenner}, \citenamefont {Siddiqi}, \citenamefont
  {Vijay}, \citenamefont {Cleland},\ and\ \citenamefont
  {Martinis}}]{Mutus2013}%
  \BibitemOpen
  \bibfield  {author} {\bibinfo {author} {\bibfnamefont {J.~Y.}\ \bibnamefont
  {Mutus}}, \bibinfo {author} {\bibfnamefont {T.~C.}\ \bibnamefont {White}},
  \bibinfo {author} {\bibfnamefont {E.}~\bibnamefont {Jeffrey}}, \bibinfo
  {author} {\bibfnamefont {D.}~\bibnamefont {Sank}}, \bibinfo {author}
  {\bibfnamefont {R.}~\bibnamefont {Barends}}, \bibinfo {author} {\bibfnamefont
  {J.}~\bibnamefont {Bochmann}}, \bibinfo {author} {\bibfnamefont
  {Y.}~\bibnamefont {Chen}}, \bibinfo {author} {\bibfnamefont {Z.}~\bibnamefont
  {Chen}}, \bibinfo {author} {\bibfnamefont {B.}~\bibnamefont {Chiaro}},
  \bibinfo {author} {\bibfnamefont {A.}~\bibnamefont {Dunsworth}}, \bibinfo
  {author} {\bibfnamefont {J.}~\bibnamefont {Kelly}}, \bibinfo {author}
  {\bibfnamefont {A.}~\bibnamefont {Megrant}}, \bibinfo {author} {\bibfnamefont
  {C.}~\bibnamefont {Neill}}, \bibinfo {author} {\bibfnamefont {P.~J.~J.}\
  \bibnamefont {O'Malley}}, \bibinfo {author} {\bibfnamefont {P.}~\bibnamefont
  {Roushan}}, \bibinfo {author} {\bibfnamefont {A.}~\bibnamefont
  {Vainsencher}}, \bibinfo {author} {\bibfnamefont {J.}~\bibnamefont {Wenner}},
  \bibinfo {author} {\bibfnamefont {I.}~\bibnamefont {Siddiqi}}, \bibinfo
  {author} {\bibfnamefont {R.}~\bibnamefont {Vijay}}, \bibinfo {author}
  {\bibfnamefont {A.~N.}\ \bibnamefont {Cleland}}, \ and\ \bibinfo {author}
  {\bibfnamefont {J.~M.}\ \bibnamefont {Martinis}},\ }\href {\doibase
  10.1063/1.4821136} {\bibfield  {journal} {\bibinfo  {journal} {Applied
  Physical Letters}\ }\textbf {\bibinfo {volume} {103}},\ \bibinfo {pages}
  {122602} (\bibinfo {year} {2013})}\BibitemShut {NoStop}%
\bibitem [{\citenamefont {L{\"{a}}hteenm{\"{a}}ki}\ \emph
  {et~al.}(2014)\citenamefont {L{\"{a}}hteenm{\"{a}}ki}, \citenamefont
  {Vesterinen}, \citenamefont {Hassel}, \citenamefont {Paraoanu}, \citenamefont
  {Sepp{\"{a}}},\ and\ \citenamefont {Hakonen}}]{Lahteenmaki2014}%
  \BibitemOpen
  \bibfield  {author} {\bibinfo {author} {\bibfnamefont {P.}~\bibnamefont
  {L{\"{a}}hteenm{\"{a}}ki}}, \bibinfo {author} {\bibfnamefont
  {V.}~\bibnamefont {Vesterinen}}, \bibinfo {author} {\bibfnamefont
  {J.}~\bibnamefont {Hassel}}, \bibinfo {author} {\bibfnamefont {G.~S.}\
  \bibnamefont {Paraoanu}}, \bibinfo {author} {\bibfnamefont {H.}~\bibnamefont
  {Sepp{\"{a}}}}, \ and\ \bibinfo {author} {\bibfnamefont {P.}~\bibnamefont
  {Hakonen}},\ }\href {\doibase 10.1007/s10909-014-1170-0} {\bibfield
  {journal} {\bibinfo  {journal} {Journal of Low Temperature Physics}\ }\textbf
  {\bibinfo {volume} {175}},\ \bibinfo {pages} {868} (\bibinfo {year}
  {2014})}\BibitemShut {NoStop}%
\bibitem [{\citenamefont {Mutus}\ \emph {et~al.}(2014)\citenamefont {Mutus},
  \citenamefont {White}, \citenamefont {Barends}, \citenamefont {Chen},
  \citenamefont {Chen}, \citenamefont {Chiaro}, \citenamefont {Dunsworth},
  \citenamefont {Jeffrey}, \citenamefont {Kelly}, \citenamefont {Megrant},
  \citenamefont {Neill}, \citenamefont {O'Malley}, \citenamefont {Roushan},
  \citenamefont {Sank}, \citenamefont {Vainsencher}, \citenamefont {Wenner},
  \citenamefont {Sundqvist}, \citenamefont {Cleland},\ and\ \citenamefont
  {Martinis}}]{Mutus2014}%
  \BibitemOpen
  \bibfield  {author} {\bibinfo {author} {\bibfnamefont {J.~Y.}\ \bibnamefont
  {Mutus}}, \bibinfo {author} {\bibfnamefont {T.~C.}\ \bibnamefont {White}},
  \bibinfo {author} {\bibfnamefont {R.}~\bibnamefont {Barends}}, \bibinfo
  {author} {\bibfnamefont {Y.}~\bibnamefont {Chen}}, \bibinfo {author}
  {\bibfnamefont {Z.}~\bibnamefont {Chen}}, \bibinfo {author} {\bibfnamefont
  {B.}~\bibnamefont {Chiaro}}, \bibinfo {author} {\bibfnamefont
  {A.}~\bibnamefont {Dunsworth}}, \bibinfo {author} {\bibfnamefont
  {E.}~\bibnamefont {Jeffrey}}, \bibinfo {author} {\bibfnamefont
  {J.}~\bibnamefont {Kelly}}, \bibinfo {author} {\bibfnamefont
  {A.}~\bibnamefont {Megrant}}, \bibinfo {author} {\bibfnamefont
  {C.}~\bibnamefont {Neill}}, \bibinfo {author} {\bibfnamefont {P.~J.~J.}\
  \bibnamefont {O'Malley}}, \bibinfo {author} {\bibfnamefont {P.}~\bibnamefont
  {Roushan}}, \bibinfo {author} {\bibfnamefont {D.}~\bibnamefont {Sank}},
  \bibinfo {author} {\bibfnamefont {A.}~\bibnamefont {Vainsencher}}, \bibinfo
  {author} {\bibfnamefont {J.}~\bibnamefont {Wenner}}, \bibinfo {author}
  {\bibfnamefont {K.~M.}\ \bibnamefont {Sundqvist}}, \bibinfo {author}
  {\bibfnamefont {A.~N.}\ \bibnamefont {Cleland}}, \ and\ \bibinfo {author}
  {\bibfnamefont {J.~M.}\ \bibnamefont {Martinis}},\ }\href {\doibase
  10.1063/1.4886408} {\bibfield  {journal} {\bibinfo  {journal} {Applied
  Physical Letters}\ }\textbf {\bibinfo {volume} {104}},\ \bibinfo {pages}
  {263513} (\bibinfo {year} {2014})}\BibitemShut {NoStop}%
\bibitem [{\citenamefont {Zhou}\ \emph {et~al.}(2014)\citenamefont {Zhou},
  \citenamefont {Schmitt}, \citenamefont {Bertet}, \citenamefont {Vion},
  \citenamefont {Wustmann}, \citenamefont {Shumeiko},\ and\ \citenamefont
  {Esteve}}]{PhysRevB.89.214517}%
  \BibitemOpen
  \bibfield  {author} {\bibinfo {author} {\bibfnamefont {X.}~\bibnamefont
  {Zhou}}, \bibinfo {author} {\bibfnamefont {V.}~\bibnamefont {Schmitt}},
  \bibinfo {author} {\bibfnamefont {P.}~\bibnamefont {Bertet}}, \bibinfo
  {author} {\bibfnamefont {D.}~\bibnamefont {Vion}}, \bibinfo {author}
  {\bibfnamefont {W.}~\bibnamefont {Wustmann}}, \bibinfo {author}
  {\bibfnamefont {V.}~\bibnamefont {Shumeiko}}, \ and\ \bibinfo {author}
  {\bibfnamefont {D.}~\bibnamefont {Esteve}},\ }\href {\doibase
  10.1103/PhysRevB.89.214517} {\bibfield  {journal} {\bibinfo  {journal} {Phys.
  Rev. B}\ }\textbf {\bibinfo {volume} {89}},\ \bibinfo {pages} {214517}
  (\bibinfo {year} {2014})}\BibitemShut {NoStop}%
\bibitem [{\citenamefont {Roy}\ \emph {et~al.}(2015)\citenamefont {Roy},
  \citenamefont {Kundu}, \citenamefont {Chand}, \citenamefont {Vadiraj},
  \citenamefont {Ranadive}, \citenamefont {Nehra}, \citenamefont {Patankar},
  \citenamefont {Aumentado}, \citenamefont {Clerk},\ and\ \citenamefont
  {Vijay}}]{Roy2015}%
  \BibitemOpen
  \bibfield  {author} {\bibinfo {author} {\bibfnamefont {T.}~\bibnamefont
  {Roy}}, \bibinfo {author} {\bibfnamefont {S.}~\bibnamefont {Kundu}}, \bibinfo
  {author} {\bibfnamefont {M.}~\bibnamefont {Chand}}, \bibinfo {author}
  {\bibfnamefont {A.~M.}\ \bibnamefont {Vadiraj}}, \bibinfo {author}
  {\bibfnamefont {A.}~\bibnamefont {Ranadive}}, \bibinfo {author}
  {\bibfnamefont {N.}~\bibnamefont {Nehra}}, \bibinfo {author} {\bibfnamefont
  {M.~P.}\ \bibnamefont {Patankar}}, \bibinfo {author} {\bibfnamefont
  {J.}~\bibnamefont {Aumentado}}, \bibinfo {author} {\bibfnamefont {A.~A.}\
  \bibnamefont {Clerk}}, \ and\ \bibinfo {author} {\bibfnamefont
  {R.}~\bibnamefont {Vijay}},\ }\href {\doibase 10.1063/1.4939148} {\bibfield
  {journal} {\bibinfo  {journal} {Applied Physics Letters}\ }\textbf {\bibinfo
  {volume} {107}},\ \bibinfo {pages} {262601} (\bibinfo {year}
  {2015})}\BibitemShut {NoStop}%
\bibitem [{\citenamefont {Jebari}\ \emph {et~al.}(2018)\citenamefont {Jebari},
  \citenamefont {Blanchet}, \citenamefont {Grimm}, \citenamefont {Hazra},
  \citenamefont {Albert}, \citenamefont {Joyez}, \citenamefont {Vion},
  \citenamefont {Est{\`{e}}ve}, \citenamefont {Portier},\ and\ \citenamefont
  {Hofheinz}}]{Jebari2018}%
  \BibitemOpen
  \bibfield  {author} {\bibinfo {author} {\bibfnamefont {S.}~\bibnamefont
  {Jebari}}, \bibinfo {author} {\bibfnamefont {F.}~\bibnamefont {Blanchet}},
  \bibinfo {author} {\bibfnamefont {A.}~\bibnamefont {Grimm}}, \bibinfo
  {author} {\bibfnamefont {D.}~\bibnamefont {Hazra}}, \bibinfo {author}
  {\bibfnamefont {R.}~\bibnamefont {Albert}}, \bibinfo {author} {\bibfnamefont
  {P.}~\bibnamefont {Joyez}}, \bibinfo {author} {\bibfnamefont
  {D.}~\bibnamefont {Vion}}, \bibinfo {author} {\bibfnamefont {D.}~\bibnamefont
  {Est{\`{e}}ve}}, \bibinfo {author} {\bibfnamefont {F.}~\bibnamefont
  {Portier}}, \ and\ \bibinfo {author} {\bibfnamefont {M.}~\bibnamefont
  {Hofheinz}},\ }\href {\doibase 10.1038/s41928-018-0055-7} {\bibfield
  {journal} {\bibinfo  {journal} {Nature Electronics}\ }\textbf {\bibinfo
  {volume} {1}},\ \bibinfo {pages} {223} (\bibinfo {year} {2018})}\BibitemShut
  {NoStop}%
\bibitem [{\citenamefont {Elo}\ \emph {et~al.}(2019)\citenamefont {Elo},
  \citenamefont {Abhilash}, \citenamefont {Perelshtein}, \citenamefont {Lilja},
  \citenamefont {Korostylev},\ and\ \citenamefont {Hakonen}}]{Elo2019}%
  \BibitemOpen
  \bibfield  {author} {\bibinfo {author} {\bibfnamefont {T.}~\bibnamefont
  {Elo}}, \bibinfo {author} {\bibfnamefont {T.~S.}\ \bibnamefont {Abhilash}},
  \bibinfo {author} {\bibfnamefont {M.~R.}\ \bibnamefont {Perelshtein}},
  \bibinfo {author} {\bibfnamefont {I.}~\bibnamefont {Lilja}}, \bibinfo
  {author} {\bibfnamefont {E.~V.}\ \bibnamefont {Korostylev}}, \ and\ \bibinfo
  {author} {\bibfnamefont {P.~J.}\ \bibnamefont {Hakonen}},\ }\href {\doibase
  10.1063/1.5086091} {\bibfield  {journal} {\bibinfo  {journal} {Applied
  Physics Letters}\ }\textbf {\bibinfo {volume} {114}},\ \bibinfo {pages}
  {152601} (\bibinfo {year} {2019})}\BibitemShut {NoStop}%
\bibitem [{\citenamefont {Eichler}\ \emph {et~al.}(2011)\citenamefont
  {Eichler}, \citenamefont {Bozyigit}, \citenamefont {Lang}, \citenamefont
  {Baur}, \citenamefont {Steffen}, \citenamefont {Fink}, \citenamefont
  {Filipp},\ and\ \citenamefont {Wallraff}}]{Eichler2011}%
  \BibitemOpen
  \bibfield  {author} {\bibinfo {author} {\bibfnamefont {C.}~\bibnamefont
  {Eichler}}, \bibinfo {author} {\bibfnamefont {D.}~\bibnamefont {Bozyigit}},
  \bibinfo {author} {\bibfnamefont {C.}~\bibnamefont {Lang}}, \bibinfo {author}
  {\bibfnamefont {M.}~\bibnamefont {Baur}}, \bibinfo {author} {\bibfnamefont
  {L.}~\bibnamefont {Steffen}}, \bibinfo {author} {\bibfnamefont {J.~M.}\
  \bibnamefont {Fink}}, \bibinfo {author} {\bibfnamefont {S.}~\bibnamefont
  {Filipp}}, \ and\ \bibinfo {author} {\bibfnamefont {A.}~\bibnamefont
  {Wallraff}},\ }\href {\doibase 10.1103/PhysRevLett.107.113601} {\bibfield
  {journal} {\bibinfo  {journal} {Physical Review Letters}\ }\textbf {\bibinfo
  {volume} {107}},\ \bibinfo {pages} {113601} (\bibinfo {year}
  {2011})}\BibitemShut {NoStop}%
\bibitem [{\citenamefont {Wilson}\ \emph {et~al.}(2011)\citenamefont {Wilson},
  \citenamefont {Johansson}, \citenamefont {Pourkabirian}, \citenamefont
  {Simoen}, \citenamefont {Johansson}, \citenamefont {Duty}, \citenamefont
  {Nori},\ and\ \citenamefont {Delsing}}]{Wilson2011}%
  \BibitemOpen
  \bibfield  {author} {\bibinfo {author} {\bibfnamefont {C.~M.}\ \bibnamefont
  {Wilson}}, \bibinfo {author} {\bibfnamefont {G.}~\bibnamefont {Johansson}},
  \bibinfo {author} {\bibfnamefont {A.}~\bibnamefont {Pourkabirian}}, \bibinfo
  {author} {\bibfnamefont {M.}~\bibnamefont {Simoen}}, \bibinfo {author}
  {\bibfnamefont {J.~R.}\ \bibnamefont {Johansson}}, \bibinfo {author}
  {\bibfnamefont {T.}~\bibnamefont {Duty}}, \bibinfo {author} {\bibfnamefont
  {F.}~\bibnamefont {Nori}}, \ and\ \bibinfo {author} {\bibfnamefont
  {P.}~\bibnamefont {Delsing}},\ }\href {\doibase 10.1038/nature10561}
  {\bibfield  {journal} {\bibinfo  {journal} {Nature}\ }\textbf {\bibinfo
  {volume} {479}},\ \bibinfo {pages} {376} (\bibinfo {year}
  {2011})}\BibitemShut {NoStop}%
\bibitem [{\citenamefont {Zhong}\ \emph {et~al.}(2013)\citenamefont {Zhong},
  \citenamefont {Menzel}, \citenamefont {Candia}, \citenamefont {Eder},
  \citenamefont {Ihmig}, \citenamefont {Baust}, \citenamefont {Haeberlein},
  \citenamefont {Hoffmann}, \citenamefont {Inomata}, \citenamefont {Yamamoto},
  \citenamefont {Nakamura}, \citenamefont {Solano}, \citenamefont {Deppe},
  \citenamefont {Marx},\ and\ \citenamefont {Gross}}]{Zhong2013}%
  \BibitemOpen
  \bibfield  {author} {\bibinfo {author} {\bibfnamefont {L.}~\bibnamefont
  {Zhong}}, \bibinfo {author} {\bibfnamefont {E.~P.}\ \bibnamefont {Menzel}},
  \bibinfo {author} {\bibfnamefont {R.~D.}\ \bibnamefont {Candia}}, \bibinfo
  {author} {\bibfnamefont {P.}~\bibnamefont {Eder}}, \bibinfo {author}
  {\bibfnamefont {M.}~\bibnamefont {Ihmig}}, \bibinfo {author} {\bibfnamefont
  {A.}~\bibnamefont {Baust}}, \bibinfo {author} {\bibfnamefont
  {M.}~\bibnamefont {Haeberlein}}, \bibinfo {author} {\bibfnamefont
  {E.}~\bibnamefont {Hoffmann}}, \bibinfo {author} {\bibfnamefont
  {K.}~\bibnamefont {Inomata}}, \bibinfo {author} {\bibfnamefont
  {T.}~\bibnamefont {Yamamoto}}, \bibinfo {author} {\bibfnamefont
  {Y.}~\bibnamefont {Nakamura}}, \bibinfo {author} {\bibfnamefont
  {E.}~\bibnamefont {Solano}}, \bibinfo {author} {\bibfnamefont
  {F.}~\bibnamefont {Deppe}}, \bibinfo {author} {\bibfnamefont
  {A.}~\bibnamefont {Marx}}, \ and\ \bibinfo {author} {\bibfnamefont
  {R.}~\bibnamefont {Gross}},\ }\href {\doibase 10.1088/1367-2630/15/12/125013}
  {\bibfield  {journal} {\bibinfo  {journal} {New Journal of Physics}\ }\textbf
  {\bibinfo {volume} {15}},\ \bibinfo {pages} {125013} (\bibinfo {year}
  {2013})}\BibitemShut {NoStop}%
\bibitem [{\citenamefont {L\"{a}hteenm\"{a}ki}\ \emph
  {et~al.}(2016)\citenamefont {L\"{a}hteenm\"{a}ki}, \citenamefont {Paraoanu},
  \citenamefont {Hassel},\ and\ \citenamefont {Hakonen}}]{Lhteenmki2016}%
  \BibitemOpen
  \bibfield  {author} {\bibinfo {author} {\bibfnamefont {P.}~\bibnamefont
  {L\"{a}hteenm\"{a}ki}}, \bibinfo {author} {\bibfnamefont {G.~S.}\
  \bibnamefont {Paraoanu}}, \bibinfo {author} {\bibfnamefont {J.}~\bibnamefont
  {Hassel}}, \ and\ \bibinfo {author} {\bibfnamefont {P.~J.}\ \bibnamefont
  {Hakonen}},\ }\href {\doibase 10.1038/ncomms12548} {\bibfield  {journal}
  {\bibinfo  {journal} {Nature Communications}\ }\textbf {\bibinfo {volume}
  {7}},\ \bibinfo {pages} {12548} (\bibinfo {year} {2016})}\BibitemShut
  {NoStop}%
\bibitem [{\citenamefont {Grimsmo}\ and\ \citenamefont
  {Blais}(2017)}]{Grimsmo2017}%
  \BibitemOpen
  \bibfield  {author} {\bibinfo {author} {\bibfnamefont {A.~L.}\ \bibnamefont
  {Grimsmo}}\ and\ \bibinfo {author} {\bibfnamefont {A.}~\bibnamefont
  {Blais}},\ }\href {\doibase 10.1038/s41534-017-0020-8} {\bibfield  {journal}
  {\bibinfo  {journal} {npj Quantum Information}\ }\textbf {\bibinfo {volume}
  {3}} (\bibinfo {year} {2017}),\ 10.1038/s41534-017-0020-8}\BibitemShut
  {NoStop}%
\bibitem [{\citenamefont {Schneider}\ \emph {et~al.}(2020)\citenamefont
  {Schneider}, \citenamefont {Bengtsson}, \citenamefont {Svensson},
  \citenamefont {Aref}, \citenamefont {Johansson}, \citenamefont {Bylander},\
  and\ \citenamefont {Delsing}}]{PhysRevLett.124.140503}%
  \BibitemOpen
  \bibfield  {author} {\bibinfo {author} {\bibfnamefont {B.~H.}\ \bibnamefont
  {Schneider}}, \bibinfo {author} {\bibfnamefont {A.}~\bibnamefont
  {Bengtsson}}, \bibinfo {author} {\bibfnamefont {I.~M.}\ \bibnamefont
  {Svensson}}, \bibinfo {author} {\bibfnamefont {T.}~\bibnamefont {Aref}},
  \bibinfo {author} {\bibfnamefont {G.}~\bibnamefont {Johansson}}, \bibinfo
  {author} {\bibfnamefont {J.}~\bibnamefont {Bylander}}, \ and\ \bibinfo
  {author} {\bibfnamefont {P.}~\bibnamefont {Delsing}},\ }\href {\doibase
  10.1103/PhysRevLett.124.140503} {\bibfield  {journal} {\bibinfo  {journal}
  {Phys. Rev. Lett.}\ }\textbf {\bibinfo {volume} {124}},\ \bibinfo {pages}
  {140503} (\bibinfo {year} {2020})}\BibitemShut {NoStop}%
\bibitem [{\citenamefont {Macklin}\ \emph {et~al.}(2015)\citenamefont
  {Macklin}, \citenamefont {O{\textquotesingle}Brien}, \citenamefont {Hover},
  \citenamefont {Schwartz}, \citenamefont {Bolkhovsky}, \citenamefont {Zhang},
  \citenamefont {Oliver},\ and\ \citenamefont {Siddiqi}}]{Macklin2015}%
  \BibitemOpen
  \bibfield  {author} {\bibinfo {author} {\bibfnamefont {C.}~\bibnamefont
  {Macklin}}, \bibinfo {author} {\bibfnamefont {K.}~\bibnamefont
  {O{\textquotesingle}Brien}}, \bibinfo {author} {\bibfnamefont
  {D.}~\bibnamefont {Hover}}, \bibinfo {author} {\bibfnamefont {M.~E.}\
  \bibnamefont {Schwartz}}, \bibinfo {author} {\bibfnamefont {V.}~\bibnamefont
  {Bolkhovsky}}, \bibinfo {author} {\bibfnamefont {X.}~\bibnamefont {Zhang}},
  \bibinfo {author} {\bibfnamefont {W.~D.}\ \bibnamefont {Oliver}}, \ and\
  \bibinfo {author} {\bibfnamefont {I.}~\bibnamefont {Siddiqi}},\ }\href
  {\doibase 10.1126/science.aaa8525} {\bibfield  {journal} {\bibinfo  {journal}
  {Science}\ }\textbf {\bibinfo {volume} {350}},\ \bibinfo {pages} {307}
  (\bibinfo {year} {2015})}\BibitemShut {NoStop}%
\bibitem [{\citenamefont {White}\ \emph {et~al.}(2015)\citenamefont {White},
  \citenamefont {Mutus}, \citenamefont {Hoi}, \citenamefont {Barends},
  \citenamefont {Campbell}, \citenamefont {Chen}, \citenamefont {Chen},
  \citenamefont {Chiaro}, \citenamefont {Dunsworth}, \citenamefont {Jeffrey},
  \citenamefont {Kelly}, \citenamefont {Megrant}, \citenamefont {Neill},
  \citenamefont {O{\textquotesingle}Malley}, \citenamefont {Roushan},
  \citenamefont {Sank}, \citenamefont {Vainsencher}, \citenamefont {Wenner},
  \citenamefont {Chaudhuri}, \citenamefont {Gao},\ and\ \citenamefont
  {Martinis}}]{White2015}%
  \BibitemOpen
  \bibfield  {author} {\bibinfo {author} {\bibfnamefont {T.~C.}\ \bibnamefont
  {White}}, \bibinfo {author} {\bibfnamefont {J.~Y.}\ \bibnamefont {Mutus}},
  \bibinfo {author} {\bibfnamefont {I.-C.}\ \bibnamefont {Hoi}}, \bibinfo
  {author} {\bibfnamefont {R.}~\bibnamefont {Barends}}, \bibinfo {author}
  {\bibfnamefont {B.}~\bibnamefont {Campbell}}, \bibinfo {author}
  {\bibfnamefont {Y.}~\bibnamefont {Chen}}, \bibinfo {author} {\bibfnamefont
  {Z.}~\bibnamefont {Chen}}, \bibinfo {author} {\bibfnamefont {B.}~\bibnamefont
  {Chiaro}}, \bibinfo {author} {\bibfnamefont {A.}~\bibnamefont {Dunsworth}},
  \bibinfo {author} {\bibfnamefont {E.}~\bibnamefont {Jeffrey}}, \bibinfo
  {author} {\bibfnamefont {J.}~\bibnamefont {Kelly}}, \bibinfo {author}
  {\bibfnamefont {A.}~\bibnamefont {Megrant}}, \bibinfo {author} {\bibfnamefont
  {C.}~\bibnamefont {Neill}}, \bibinfo {author} {\bibfnamefont {P.~J.~J.}\
  \bibnamefont {O{\textquotesingle}Malley}}, \bibinfo {author} {\bibfnamefont
  {P.}~\bibnamefont {Roushan}}, \bibinfo {author} {\bibfnamefont
  {D.}~\bibnamefont {Sank}}, \bibinfo {author} {\bibfnamefont {A.}~\bibnamefont
  {Vainsencher}}, \bibinfo {author} {\bibfnamefont {J.}~\bibnamefont {Wenner}},
  \bibinfo {author} {\bibfnamefont {S.}~\bibnamefont {Chaudhuri}}, \bibinfo
  {author} {\bibfnamefont {J.}~\bibnamefont {Gao}}, \ and\ \bibinfo {author}
  {\bibfnamefont {J.~M.}\ \bibnamefont {Martinis}},\ }\href {\doibase
  10.1063/1.4922348} {\bibfield  {journal} {\bibinfo  {journal} {Applied
  Physics Letters}\ }\textbf {\bibinfo {volume} {106}},\ \bibinfo {pages}
  {242601} (\bibinfo {year} {2015})}\BibitemShut {NoStop}%
\bibitem [{\citenamefont {Zorin}(2016)}]{Zorin2016}%
  \BibitemOpen
  \bibfield  {author} {\bibinfo {author} {\bibfnamefont {A.~B.}\ \bibnamefont
  {Zorin}},\ }\href {\doibase 10.1103/PhysRevApplied.6.034006} {\bibfield
  {journal} {\bibinfo  {journal} {Phys. Rev. Applied}\ }\textbf {\bibinfo
  {volume} {6}},\ \bibinfo {pages} {034006} (\bibinfo {year}
  {2016})}\BibitemShut {NoStop}%
\bibitem [{\citenamefont {Krinner}\ \emph {et~al.}(2019)\citenamefont
  {Krinner}, \citenamefont {Storz}, \citenamefont {Kurpiers}, \citenamefont
  {Magnard}, \citenamefont {Heinsoo}, \citenamefont {Keller}, \citenamefont
  {L\"{u}tolf}, \citenamefont {Eichler},\ and\ \citenamefont
  {Wallraff}}]{Krinner2019}%
  \BibitemOpen
  \bibfield  {author} {\bibinfo {author} {\bibfnamefont {S.}~\bibnamefont
  {Krinner}}, \bibinfo {author} {\bibfnamefont {S.}~\bibnamefont {Storz}},
  \bibinfo {author} {\bibfnamefont {P.}~\bibnamefont {Kurpiers}}, \bibinfo
  {author} {\bibfnamefont {P.}~\bibnamefont {Magnard}}, \bibinfo {author}
  {\bibfnamefont {J.}~\bibnamefont {Heinsoo}}, \bibinfo {author} {\bibfnamefont
  {R.}~\bibnamefont {Keller}}, \bibinfo {author} {\bibfnamefont
  {J.}~\bibnamefont {L\"{u}tolf}}, \bibinfo {author} {\bibfnamefont
  {C.}~\bibnamefont {Eichler}}, \ and\ \bibinfo {author} {\bibfnamefont
  {A.}~\bibnamefont {Wallraff}},\ }\href {\doibase
  10.1140/epjqt/s40507-019-0072-0} {\bibfield  {journal} {\bibinfo  {journal}
  {{EPJ} Quantum Technology}\ }\textbf {\bibinfo {volume} {6}},\ \bibinfo
  {pages} {2} (\bibinfo {year} {2019})}\BibitemShut {NoStop}%
\bibitem [{\citenamefont {Cai}\ \emph {et~al.}(2017)\citenamefont {Cai},
  \citenamefont {Roslund}, \citenamefont {Ferrini}, \citenamefont {Arzani},
  \citenamefont {Xu}, \citenamefont {Fabre},\ and\ \citenamefont
  {Treps}}]{Cai2017}%
  \BibitemOpen
  \bibfield  {author} {\bibinfo {author} {\bibfnamefont {Y.}~\bibnamefont
  {Cai}}, \bibinfo {author} {\bibfnamefont {J.}~\bibnamefont {Roslund}},
  \bibinfo {author} {\bibfnamefont {G.}~\bibnamefont {Ferrini}}, \bibinfo
  {author} {\bibfnamefont {F.}~\bibnamefont {Arzani}}, \bibinfo {author}
  {\bibfnamefont {X.}~\bibnamefont {Xu}}, \bibinfo {author} {\bibfnamefont
  {C.}~\bibnamefont {Fabre}}, \ and\ \bibinfo {author} {\bibfnamefont
  {N.}~\bibnamefont {Treps}},\ }\href {https://doi.org/10.1038/ncomms15645}
  {\bibfield  {journal} {\bibinfo  {journal} {Nature Communications}\ }\textbf
  {\bibinfo {volume} {8}} (\bibinfo {year} {2017})}\BibitemShut {NoStop}%
\bibitem [{\citenamefont {Esposito}\ \emph
  {et~al.}(2021{\natexlab{a}})\citenamefont {Esposito}, \citenamefont
  {Ranadive}, \citenamefont {Planat},\ and\ \citenamefont
  {Roch}}]{Esposito2021}%
  \BibitemOpen
  \bibfield  {author} {\bibinfo {author} {\bibfnamefont {M.}~\bibnamefont
  {Esposito}}, \bibinfo {author} {\bibfnamefont {A.}~\bibnamefont {Ranadive}},
  \bibinfo {author} {\bibfnamefont {L.}~\bibnamefont {Planat}}, \ and\ \bibinfo
  {author} {\bibfnamefont {N.}~\bibnamefont {Roch}},\ }\href {\doibase
  10.1063/5.0064892} {\bibfield  {journal} {\bibinfo  {journal} {Applied
  Physics Letters}\ }\textbf {\bibinfo {volume} {119}},\ \bibinfo {pages}
  {120501} (\bibinfo {year} {2021}{\natexlab{a}})}\BibitemShut {NoStop}%
\bibitem [{\citenamefont {Vesterinen}\ and\ \citenamefont
  {Hassel}(2020)}]{patent}%
  \BibitemOpen
  \bibfield  {author} {\bibinfo {author} {\bibfnamefont {V.}~\bibnamefont
  {Vesterinen}}\ and\ \bibinfo {author} {\bibfnamefont {J.}~\bibnamefont
  {Hassel}},\ }\href
  {https://worldwide.espacenet.com/patent/search/family/069174515/publication/FI20195045A1?q=pn\%3DFI20195045A1}
  {\enquote {\bibinfo {title} {{J}osephson traveling wave parametric amplifier,
  {WO} 2020/152393 {A}1},}\ } (\bibinfo {year} {2020})\BibitemShut {NoStop}%
\bibitem [{\citenamefont {Frattini}\ \emph {et~al.}(2017)\citenamefont
  {Frattini}, \citenamefont {Vool}, \citenamefont {Shankar}, \citenamefont
  {Narla}, \citenamefont {Sliwa},\ and\ \citenamefont
  {Devoret}}]{Frattini2017}%
  \BibitemOpen
  \bibfield  {author} {\bibinfo {author} {\bibfnamefont {N.~E.}\ \bibnamefont
  {Frattini}}, \bibinfo {author} {\bibfnamefont {U.}~\bibnamefont {Vool}},
  \bibinfo {author} {\bibfnamefont {S.}~\bibnamefont {Shankar}}, \bibinfo
  {author} {\bibfnamefont {A.}~\bibnamefont {Narla}}, \bibinfo {author}
  {\bibfnamefont {K.~M.}\ \bibnamefont {Sliwa}}, \ and\ \bibinfo {author}
  {\bibfnamefont {M.~H.}\ \bibnamefont {Devoret}},\ }\href {\doibase
  10.1063/1.4984142} {\bibfield  {journal} {\bibinfo  {journal} {Applied
  Physics Letters}\ }\textbf {\bibinfo {volume} {110}},\ \bibinfo {pages}
  {222603} (\bibinfo {year} {2017})}\BibitemShut {NoStop}%
\bibitem [{\citenamefont {Yurke}\ and\ \citenamefont {Buks}(2006)}]{Yurke2006}%
  \BibitemOpen
  \bibfield  {author} {\bibinfo {author} {\bibfnamefont {B.}~\bibnamefont
  {Yurke}}\ and\ \bibinfo {author} {\bibfnamefont {E.}~\bibnamefont {Buks}},\
  }\href {\doibase 10.1109/jlt.2006.884490} {\bibfield  {journal} {\bibinfo
  {journal} {Journal of Lightwave Technology}\ }\textbf {\bibinfo {volume}
  {24}},\ \bibinfo {pages} {5054} (\bibinfo {year} {2006})}\BibitemShut
  {NoStop}%
\bibitem [{\citenamefont {Krupko}\ \emph {et~al.}(2018)\citenamefont {Krupko},
  \citenamefont {Nguyen}, \citenamefont {Wei{\ss}l}, \citenamefont {Dumur},
  \citenamefont {Puertas}, \citenamefont {Dassonneville}, \citenamefont {Naud},
  \citenamefont {Hekking}, \citenamefont {Basko}, \citenamefont {Buisson},
  \citenamefont {Roch},\ and\ \citenamefont {Hasch-Guichard}}]{Krupko2018}%
  \BibitemOpen
  \bibfield  {author} {\bibinfo {author} {\bibfnamefont {Y.}~\bibnamefont
  {Krupko}}, \bibinfo {author} {\bibfnamefont {V.~D.}\ \bibnamefont {Nguyen}},
  \bibinfo {author} {\bibfnamefont {T.}~\bibnamefont {Wei{\ss}l}}, \bibinfo
  {author} {\bibfnamefont {{\'{E}}.}~\bibnamefont {Dumur}}, \bibinfo {author}
  {\bibfnamefont {J.}~\bibnamefont {Puertas}}, \bibinfo {author} {\bibfnamefont
  {R.}~\bibnamefont {Dassonneville}}, \bibinfo {author} {\bibfnamefont
  {C.}~\bibnamefont {Naud}}, \bibinfo {author} {\bibfnamefont {F.~W.~J.}\
  \bibnamefont {Hekking}}, \bibinfo {author} {\bibfnamefont {D.~M.}\
  \bibnamefont {Basko}}, \bibinfo {author} {\bibfnamefont {O.}~\bibnamefont
  {Buisson}}, \bibinfo {author} {\bibfnamefont {N.}~\bibnamefont {Roch}}, \
  and\ \bibinfo {author} {\bibfnamefont {W.}~\bibnamefont {Hasch-Guichard}},\
  }\href {\doibase 10.1103/physrevb.98.094516} {\bibfield  {journal} {\bibinfo
  {journal} {Physical Review B}\ }\textbf {\bibinfo {volume} {98}},\ \bibinfo
  {pages} {094516} (\bibinfo {year} {2018})}\BibitemShut {NoStop}%
\bibitem [{\citenamefont {Planat}\ \emph {et~al.}(2020)\citenamefont {Planat},
  \citenamefont {Ranadive}, \citenamefont {Dassonneville}, \citenamefont
  {Puertas~Mart\'{\i}nez}, \citenamefont {L\'eger}, \citenamefont {Naud},
  \citenamefont {Buisson}, \citenamefont {Hasch-Guichard}, \citenamefont
  {Basko},\ and\ \citenamefont {Roch}}]{Planat2020}%
  \BibitemOpen
  \bibfield  {author} {\bibinfo {author} {\bibfnamefont {L.}~\bibnamefont
  {Planat}}, \bibinfo {author} {\bibfnamefont {A.}~\bibnamefont {Ranadive}},
  \bibinfo {author} {\bibfnamefont {R.}~\bibnamefont {Dassonneville}}, \bibinfo
  {author} {\bibfnamefont {J.}~\bibnamefont {Puertas~Mart\'{\i}nez}}, \bibinfo
  {author} {\bibfnamefont {S.}~\bibnamefont {L\'eger}}, \bibinfo {author}
  {\bibfnamefont {C.}~\bibnamefont {Naud}}, \bibinfo {author} {\bibfnamefont
  {O.}~\bibnamefont {Buisson}}, \bibinfo {author} {\bibfnamefont
  {W.}~\bibnamefont {Hasch-Guichard}}, \bibinfo {author} {\bibfnamefont
  {D.~M.}\ \bibnamefont {Basko}}, \ and\ \bibinfo {author} {\bibfnamefont
  {N.}~\bibnamefont {Roch}},\ }\href {\doibase 10.1103/PhysRevX.10.021021}
  {\bibfield  {journal} {\bibinfo  {journal} {Physical Review X}\ }\textbf
  {\bibinfo {volume} {10}},\ \bibinfo {pages} {021021} (\bibinfo {year}
  {2020})}\BibitemShut {NoStop}%
\bibitem [{\citenamefont {Agrawal}(2013)}]{Agrawal2013}%
  \BibitemOpen
  \bibfield  {author} {\bibinfo {author} {\bibfnamefont {G.}~\bibnamefont
  {Agrawal}},\ }\href {\doibase 10.1016/c2011-0-00045-5} {\emph {\bibinfo
  {title} {Nonlinear Fiber Optics}}}\ (\bibinfo  {publisher} {Elsevier},\
  \bibinfo {year} {2013})\BibitemShut {NoStop}%
\bibitem [{\citenamefont {Malnou}\ \emph {et~al.}(2021)\citenamefont {Malnou},
  \citenamefont {Vissers}, \citenamefont {Wheeler}, \citenamefont {Aumentado},
  \citenamefont {Hubmayr}, \citenamefont {Ullom},\ and\ \citenamefont
  {Gao}}]{Malnou2021}%
  \BibitemOpen
  \bibfield  {author} {\bibinfo {author} {\bibfnamefont {M.}~\bibnamefont
  {Malnou}}, \bibinfo {author} {\bibfnamefont {M.}~\bibnamefont {Vissers}},
  \bibinfo {author} {\bibfnamefont {J.}~\bibnamefont {Wheeler}}, \bibinfo
  {author} {\bibfnamefont {J.}~\bibnamefont {Aumentado}}, \bibinfo {author}
  {\bibfnamefont {J.}~\bibnamefont {Hubmayr}}, \bibinfo {author} {\bibfnamefont
  {J.}~\bibnamefont {Ullom}}, \ and\ \bibinfo {author} {\bibfnamefont
  {J.}~\bibnamefont {Gao}},\ }\href {\doibase 10.1103/PRXQuantum.2.010302}
  {\bibfield  {journal} {\bibinfo  {journal} {PRX Quantum}\ }\textbf {\bibinfo
  {volume} {2}},\ \bibinfo {pages} {010302} (\bibinfo {year}
  {2021})}\BibitemShut {NoStop}%
\bibitem [{\citenamefont {Ranadive}\ \emph {et~al.}(2021)\citenamefont
  {Ranadive}, \citenamefont {Esposito}, \citenamefont {Planat}, \citenamefont
  {Bonet}, \citenamefont {Naud}, \citenamefont {Buisson}, \citenamefont
  {Guichard},\ and\ \citenamefont {Roch}}]{Ranadive2021}%
  \BibitemOpen
  \bibfield  {author} {\bibinfo {author} {\bibfnamefont {A.}~\bibnamefont
  {Ranadive}}, \bibinfo {author} {\bibfnamefont {M.}~\bibnamefont {Esposito}},
  \bibinfo {author} {\bibfnamefont {L.}~\bibnamefont {Planat}}, \bibinfo
  {author} {\bibfnamefont {E.}~\bibnamefont {Bonet}}, \bibinfo {author}
  {\bibfnamefont {C.}~\bibnamefont {Naud}}, \bibinfo {author} {\bibfnamefont
  {O.}~\bibnamefont {Buisson}}, \bibinfo {author} {\bibfnamefont
  {W.}~\bibnamefont {Guichard}}, \ and\ \bibinfo {author} {\bibfnamefont
  {N.}~\bibnamefont {Roch}},\ }\href@noop {} {\enquote {\bibinfo {title} {A
  reversed {K}err traveling wave parametric amplifier},}\ } (\bibinfo {year}
  {2021}),\ \Eprint {http://arxiv.org/abs/2101.05815} {arXiv:2101.05815
  [quant-ph]} \BibitemShut {NoStop}%
\bibitem [{\citenamefont {Bell}\ and\ \citenamefont
  {Samolov}(2015)}]{Bell2015}%
  \BibitemOpen
  \bibfield  {author} {\bibinfo {author} {\bibfnamefont {M.~T.}\ \bibnamefont
  {Bell}}\ and\ \bibinfo {author} {\bibfnamefont {A.}~\bibnamefont {Samolov}},\
  }\href {\doibase 10.1103/PhysRevApplied.4.024014} {\bibfield  {journal}
  {\bibinfo  {journal} {Phys. Rev. Applied}\ }\textbf {\bibinfo {volume} {4}},\
  \bibinfo {pages} {024014} (\bibinfo {year} {2015})}\BibitemShut {NoStop}%
\bibitem [{\citenamefont {Boutin}\ \emph {et~al.}(2017)\citenamefont {Boutin},
  \citenamefont {Toyli}, \citenamefont {Venkatramani}, \citenamefont {Eddins},
  \citenamefont {Siddiqi},\ and\ \citenamefont {Blais}}]{Boutin2017}%
  \BibitemOpen
  \bibfield  {author} {\bibinfo {author} {\bibfnamefont {S.}~\bibnamefont
  {Boutin}}, \bibinfo {author} {\bibfnamefont {D.~M.}\ \bibnamefont {Toyli}},
  \bibinfo {author} {\bibfnamefont {A.~V.}\ \bibnamefont {Venkatramani}},
  \bibinfo {author} {\bibfnamefont {A.~W.}\ \bibnamefont {Eddins}}, \bibinfo
  {author} {\bibfnamefont {I.}~\bibnamefont {Siddiqi}}, \ and\ \bibinfo
  {author} {\bibfnamefont {A.}~\bibnamefont {Blais}},\ }\href {\doibase
  10.1103/PhysRevApplied.8.054030} {\bibfield  {journal} {\bibinfo  {journal}
  {Physical Review Applied}\ }\textbf {\bibinfo {volume} {8}},\ \bibinfo
  {pages} {054030} (\bibinfo {year} {2017})}\BibitemShut {NoStop}%
\bibitem [{\citenamefont {Peng}\ \emph {et~al.}(2021)\citenamefont {Peng},
  \citenamefont {Naghiloo}, \citenamefont {Wang}, \citenamefont {Cunningham},
  \citenamefont {Ye},\ and\ \citenamefont {O'Brien}}]{Peng2021}%
  \BibitemOpen
  \bibfield  {author} {\bibinfo {author} {\bibfnamefont {K.}~\bibnamefont
  {Peng}}, \bibinfo {author} {\bibfnamefont {M.}~\bibnamefont {Naghiloo}},
  \bibinfo {author} {\bibfnamefont {J.}~\bibnamefont {Wang}}, \bibinfo {author}
  {\bibfnamefont {G.~D.}\ \bibnamefont {Cunningham}}, \bibinfo {author}
  {\bibfnamefont {Y.}~\bibnamefont {Ye}}, \ and\ \bibinfo {author}
  {\bibfnamefont {K.~P.}\ \bibnamefont {O'Brien}},\ }\href@noop {} {\enquote
  {\bibinfo {title} {Near-ideal quantum efficiency with a {F}loquet mode
  traveling wave parametric amplfier},}\ } (\bibinfo {year} {2021}),\ \Eprint
  {http://arxiv.org/abs/arXiv 2104.08269} {arXiv:arXiv 2104.08269 [quant-ph]}
  \BibitemShut {NoStop}%
\bibitem [{SI()}]{SI}%
  \BibitemOpen
  \href@noop {} {\enquote {\bibinfo {title} {See {S}upplemental {M}aterial for
  the experimental details, loss model, and noise calibration.}}\ }\BibitemShut
  {NoStop}%
\bibitem [{\citenamefont {Grönberg}\ \emph {et~al.}(2017)\citenamefont
  {Grönberg}, \citenamefont {Kiviranta}, \citenamefont {Vesterinen},
  \citenamefont {Lehtinen}, \citenamefont {Simbierowicz}, \citenamefont
  {Luomahaara}, \citenamefont {Prunnila},\ and\ \citenamefont
  {Hassel}}]{Gr_nberg_2017}%
  \BibitemOpen
  \bibfield  {author} {\bibinfo {author} {\bibfnamefont {L.}~\bibnamefont
  {Grönberg}}, \bibinfo {author} {\bibfnamefont {M.}~\bibnamefont
  {Kiviranta}}, \bibinfo {author} {\bibfnamefont {V.}~\bibnamefont
  {Vesterinen}}, \bibinfo {author} {\bibfnamefont {J.}~\bibnamefont
  {Lehtinen}}, \bibinfo {author} {\bibfnamefont {S.}~\bibnamefont
  {Simbierowicz}}, \bibinfo {author} {\bibfnamefont {J.}~\bibnamefont
  {Luomahaara}}, \bibinfo {author} {\bibfnamefont {M.}~\bibnamefont
  {Prunnila}}, \ and\ \bibinfo {author} {\bibfnamefont {J.}~\bibnamefont
  {Hassel}},\ }\href {\doibase 10.1088/1361-6668/aa9411} {\bibfield  {journal}
  {\bibinfo  {journal} {Superconductor Science and Technology}\ }\textbf
  {\bibinfo {volume} {30}},\ \bibinfo {pages} {125016} (\bibinfo {year}
  {2017})}\BibitemShut {NoStop}%
\bibitem [{\citenamefont {Simbierowicz}\ \emph {et~al.}(2021)\citenamefont
  {Simbierowicz}, \citenamefont {Vesterinen}, \citenamefont {Milem},
  \citenamefont {Lintunen}, \citenamefont {Oksanen}, \citenamefont {Roschier},
  \citenamefont {Gr\"{o}nberg}, \citenamefont {Hassel}, \citenamefont
  {Gunnarsson},\ and\ \citenamefont {Lake}}]{Simbierowicz2021}%
  \BibitemOpen
  \bibfield  {author} {\bibinfo {author} {\bibfnamefont {S.}~\bibnamefont
  {Simbierowicz}}, \bibinfo {author} {\bibfnamefont {V.}~\bibnamefont
  {Vesterinen}}, \bibinfo {author} {\bibfnamefont {J.}~\bibnamefont {Milem}},
  \bibinfo {author} {\bibfnamefont {A.}~\bibnamefont {Lintunen}}, \bibinfo
  {author} {\bibfnamefont {M.}~\bibnamefont {Oksanen}}, \bibinfo {author}
  {\bibfnamefont {L.}~\bibnamefont {Roschier}}, \bibinfo {author}
  {\bibfnamefont {L.}~\bibnamefont {Gr\"{o}nberg}}, \bibinfo {author}
  {\bibfnamefont {J.}~\bibnamefont {Hassel}}, \bibinfo {author} {\bibfnamefont
  {D.}~\bibnamefont {Gunnarsson}}, \ and\ \bibinfo {author} {\bibfnamefont
  {R.~E.}\ \bibnamefont {Lake}},\ }\href {\doibase 10.1063/5.0028951}
  {\bibfield  {journal} {\bibinfo  {journal} {Review of Scientific
  Instruments}\ }\textbf {\bibinfo {volume} {92}},\ \bibinfo {pages} {034708}
  (\bibinfo {year} {2021})}\BibitemShut {NoStop}%
\bibitem [{\citenamefont {Parameswaran}\ \emph {et~al.}(2002)\citenamefont
  {Parameswaran}, \citenamefont {Kurz}, \citenamefont {Roussev},\ and\
  \citenamefont {Fejer}}]{Parameswaran2002}%
  \BibitemOpen
  \bibfield  {author} {\bibinfo {author} {\bibfnamefont {K.~R.}\ \bibnamefont
  {Parameswaran}}, \bibinfo {author} {\bibfnamefont {J.~R.}\ \bibnamefont
  {Kurz}}, \bibinfo {author} {\bibfnamefont {R.~V.}\ \bibnamefont {Roussev}}, \
  and\ \bibinfo {author} {\bibfnamefont {M.~M.}\ \bibnamefont {Fejer}},\ }\href
  {\doibase 10.1364/ol.27.000043} {\bibfield  {journal} {\bibinfo  {journal}
  {Optics Letters}\ }\textbf {\bibinfo {volume} {27}},\ \bibinfo {pages} {43}
  (\bibinfo {year} {2002})}\BibitemShut {NoStop}%
\bibitem [{\citenamefont {Vesterinen}\ and\ \citenamefont
  {Simbierowicz}(2021)}]{patent2}%
  \BibitemOpen
  \bibfield  {author} {\bibinfo {author} {\bibfnamefont {V.}~\bibnamefont
  {Vesterinen}}\ and\ \bibinfo {author} {\bibfnamefont {S.}~\bibnamefont
  {Simbierowicz}},\ }\href
  {https://worldwide.espacenet.com/patent/search/family/075746653/publication/FI20205401A1?q=pn\%3DFI20205401A1}
  {\enquote {\bibinfo {title} {Traveling wave parametric amplifier
  {PCT/FI}2021/050283},}\ } (\bibinfo {year} {2021})\BibitemShut {NoStop}%
\bibitem [{\citenamefont {Dixon}\ \emph {et~al.}(2020)\citenamefont {Dixon},
  \citenamefont {Dunstan}, \citenamefont {Long}, \citenamefont {Williams},
  \citenamefont {Meeson},\ and\ \citenamefont
  {Shelly}}]{PhysRevApplied.14.034058}%
  \BibitemOpen
  \bibfield  {author} {\bibinfo {author} {\bibfnamefont {T.}~\bibnamefont
  {Dixon}}, \bibinfo {author} {\bibfnamefont {J.}~\bibnamefont {Dunstan}},
  \bibinfo {author} {\bibfnamefont {G.}~\bibnamefont {Long}}, \bibinfo {author}
  {\bibfnamefont {J.}~\bibnamefont {Williams}}, \bibinfo {author}
  {\bibfnamefont {P.}~\bibnamefont {Meeson}}, \ and\ \bibinfo {author}
  {\bibfnamefont {C.}~\bibnamefont {Shelly}},\ }\href {\doibase
  10.1103/PhysRevApplied.14.034058} {\bibfield  {journal} {\bibinfo  {journal}
  {Phys. Rev. Applied}\ }\textbf {\bibinfo {volume} {14}},\ \bibinfo {pages}
  {034058} (\bibinfo {year} {2020})}\BibitemShut {NoStop}%
\bibitem [{\citenamefont {DeSalvo}\ \emph {et~al.}(1992)\citenamefont
  {DeSalvo}, \citenamefont {Vanherzeele}, \citenamefont {Hagan}, \citenamefont
  {Sheik-Bahae}, \citenamefont {Stegeman},\ and\ \citenamefont
  {Stryland}}]{DeSalvo1992}%
  \BibitemOpen
  \bibfield  {author} {\bibinfo {author} {\bibfnamefont {R.}~\bibnamefont
  {DeSalvo}}, \bibinfo {author} {\bibfnamefont {H.}~\bibnamefont
  {Vanherzeele}}, \bibinfo {author} {\bibfnamefont {D.~J.}\ \bibnamefont
  {Hagan}}, \bibinfo {author} {\bibfnamefont {M.}~\bibnamefont {Sheik-Bahae}},
  \bibinfo {author} {\bibfnamefont {G.}~\bibnamefont {Stegeman}}, \ and\
  \bibinfo {author} {\bibfnamefont {E.~W.~V.}\ \bibnamefont {Stryland}},\
  }\href {\doibase 10.1364/ol.17.000028} {\bibfield  {journal} {\bibinfo
  {journal} {Optics Letters}\ }\textbf {\bibinfo {volume} {17}},\ \bibinfo
  {pages} {28} (\bibinfo {year} {1992})}\BibitemShut {NoStop}%
\bibitem [{\citenamefont {Bultink}\ \emph {et~al.}(2018)\citenamefont
  {Bultink}, \citenamefont {Tarasinski}, \citenamefont {Haandb{\ae}k},
  \citenamefont {Poletto}, \citenamefont {Haider}, \citenamefont {Michalak},
  \citenamefont {Bruno},\ and\ \citenamefont {DiCarlo}}]{Bultink2018}%
  \BibitemOpen
  \bibfield  {author} {\bibinfo {author} {\bibfnamefont {C.~C.}\ \bibnamefont
  {Bultink}}, \bibinfo {author} {\bibfnamefont {B.}~\bibnamefont {Tarasinski}},
  \bibinfo {author} {\bibfnamefont {N.}~\bibnamefont {Haandb{\ae}k}}, \bibinfo
  {author} {\bibfnamefont {S.}~\bibnamefont {Poletto}}, \bibinfo {author}
  {\bibfnamefont {N.}~\bibnamefont {Haider}}, \bibinfo {author} {\bibfnamefont
  {D.~J.}\ \bibnamefont {Michalak}}, \bibinfo {author} {\bibfnamefont
  {A.}~\bibnamefont {Bruno}}, \ and\ \bibinfo {author} {\bibfnamefont
  {L.}~\bibnamefont {DiCarlo}},\ }\href {\doibase 10.1063/1.5015954} {\bibfield
   {journal} {\bibinfo  {journal} {Applied Physics Letters}\ }\textbf {\bibinfo
  {volume} {112}},\ \bibinfo {pages} {092601} (\bibinfo {year}
  {2018})}\BibitemShut {NoStop}%
\bibitem [{\citenamefont {Braunstein}\ and\ \citenamefont {van
  Loock}(2005)}]{Braunstein2005}%
  \BibitemOpen
  \bibfield  {author} {\bibinfo {author} {\bibfnamefont {S.~L.}\ \bibnamefont
  {Braunstein}}\ and\ \bibinfo {author} {\bibfnamefont {P.}~\bibnamefont {van
  Loock}},\ }\href {\doibase 10.1103/revmodphys.77.513} {\bibfield  {journal}
  {\bibinfo  {journal} {Reviews of Modern Physics}\ }\textbf {\bibinfo {volume}
  {77}},\ \bibinfo {pages} {513} (\bibinfo {year} {2005})}\BibitemShut
  {NoStop}%
\bibitem [{\citenamefont {Adesso}\ \emph {et~al.}(2014)\citenamefont {Adesso},
  \citenamefont {Ragy},\ and\ \citenamefont {Lee}}]{Adesso2014}%
  \BibitemOpen
  \bibfield  {author} {\bibinfo {author} {\bibfnamefont {G.}~\bibnamefont
  {Adesso}}, \bibinfo {author} {\bibfnamefont {S.}~\bibnamefont {Ragy}}, \ and\
  \bibinfo {author} {\bibfnamefont {A.~R.}\ \bibnamefont {Lee}},\ }\href
  {\doibase 10.1142/s1230161214400010} {\bibfield  {journal} {\bibinfo
  {journal} {Open Systems {\&} Information Dynamics}\ }\textbf {\bibinfo
  {volume} {21}},\ \bibinfo {pages} {1440001} (\bibinfo {year}
  {2014})}\BibitemShut {NoStop}%
\bibitem [{\citenamefont {Peres}(1996)}]{Peres1996}%
  \BibitemOpen
  \bibfield  {author} {\bibinfo {author} {\bibfnamefont {A.}~\bibnamefont
  {Peres}},\ }\href {\doibase 10.1103/PhysRevLett.77.1413} {\bibfield
  {journal} {\bibinfo  {journal} {Physical Review Letters}\ }\textbf {\bibinfo
  {volume} {77}},\ \bibinfo {pages} {1413} (\bibinfo {year}
  {1996})}\BibitemShut {NoStop}%
\bibitem [{\citenamefont {Horodecki}(1997)}]{Horodecki1997}%
  \BibitemOpen
  \bibfield  {author} {\bibinfo {author} {\bibfnamefont {P.}~\bibnamefont
  {Horodecki}},\ }\href {\doibase 10.1016/s0375-9601(97)00416-7} {\bibfield
  {journal} {\bibinfo  {journal} {Physics Letters A}\ }\textbf {\bibinfo
  {volume} {232}},\ \bibinfo {pages} {333} (\bibinfo {year}
  {1997})}\BibitemShut {NoStop}%
\bibitem [{\citenamefont {Simon}(2000)}]{Simon2000}%
  \BibitemOpen
  \bibfield  {author} {\bibinfo {author} {\bibfnamefont {R.}~\bibnamefont
  {Simon}},\ }\href {\doibase 10.1103/PhysRevLett.84.2726} {\bibfield
  {journal} {\bibinfo  {journal} {Physical Review Letters}\ }\textbf {\bibinfo
  {volume} {84}},\ \bibinfo {pages} {2726} (\bibinfo {year}
  {2000})}\BibitemShut {NoStop}%
\bibitem [{\citenamefont {Vidal}\ and\ \citenamefont
  {Werner}(2002)}]{Vidal2002}%
  \BibitemOpen
  \bibfield  {author} {\bibinfo {author} {\bibfnamefont {G.}~\bibnamefont
  {Vidal}}\ and\ \bibinfo {author} {\bibfnamefont {R.~F.}\ \bibnamefont
  {Werner}},\ }\href {\doibase 10.1103/PhysRevA.65.032314} {\bibfield
  {journal} {\bibinfo  {journal} {Physical Review A}\ }\textbf {\bibinfo
  {volume} {65}},\ \bibinfo {pages} {032314} (\bibinfo {year}
  {2002})}\BibitemShut {NoStop}%
\bibitem [{\citenamefont {Audenaert}\ \emph {et~al.}(2003)\citenamefont
  {Audenaert}, \citenamefont {Plenio},\ and\ \citenamefont
  {Eisert}}]{Audenaert2003}%
  \BibitemOpen
  \bibfield  {author} {\bibinfo {author} {\bibfnamefont {K.}~\bibnamefont
  {Audenaert}}, \bibinfo {author} {\bibfnamefont {M.~B.}\ \bibnamefont
  {Plenio}}, \ and\ \bibinfo {author} {\bibfnamefont {J.}~\bibnamefont
  {Eisert}},\ }\href {\doibase 10.1103/PhysRevLett.90.027901} {\bibfield
  {journal} {\bibinfo  {journal} {Physical Review Letters}\ }\textbf {\bibinfo
  {volume} {90}},\ \bibinfo {pages} {027901} (\bibinfo {year}
  {2003})}\BibitemShut {NoStop}%
\bibitem [{\citenamefont {Sandbo~Chang}\ \emph {et~al.}(2018)\citenamefont
  {Sandbo~Chang}, \citenamefont {Simoen}, \citenamefont {Aumentado},
  \citenamefont {Sab\'{\i}n}, \citenamefont {Forn-D\'{\i}az}, \citenamefont
  {Vadiraj}, \citenamefont {Quijandr\'{\i}a}, \citenamefont {Johansson},
  \citenamefont {Fuentes},\ and\ \citenamefont {Wilson}}]{SandboChang2018}%
  \BibitemOpen
  \bibfield  {author} {\bibinfo {author} {\bibfnamefont {C.~W.}\ \bibnamefont
  {Sandbo~Chang}}, \bibinfo {author} {\bibfnamefont {M.}~\bibnamefont
  {Simoen}}, \bibinfo {author} {\bibfnamefont {J.}~\bibnamefont {Aumentado}},
  \bibinfo {author} {\bibfnamefont {C.}~\bibnamefont {Sab\'{\i}n}}, \bibinfo
  {author} {\bibfnamefont {P.}~\bibnamefont {Forn-D\'{\i}az}}, \bibinfo
  {author} {\bibfnamefont {A.~M.}\ \bibnamefont {Vadiraj}}, \bibinfo {author}
  {\bibfnamefont {F.}~\bibnamefont {Quijandr\'{\i}a}}, \bibinfo {author}
  {\bibfnamefont {G.}~\bibnamefont {Johansson}}, \bibinfo {author}
  {\bibfnamefont {I.}~\bibnamefont {Fuentes}}, \ and\ \bibinfo {author}
  {\bibfnamefont {C.~M.}\ \bibnamefont {Wilson}},\ }\href {\doibase
  10.1103/PhysRevApplied.10.044019} {\bibfield  {journal} {\bibinfo  {journal}
  {Physical Review Applied}\ }\textbf {\bibinfo {volume} {10}},\ \bibinfo
  {pages} {044019} (\bibinfo {year} {2018})}\BibitemShut {NoStop}%
\bibitem [{\citenamefont {Paris}\ \emph {et~al.}(2003)\citenamefont {Paris},
  \citenamefont {Illuminati}, \citenamefont {Serafini},\ and\ \citenamefont
  {De~Siena}}]{Paris2003}%
  \BibitemOpen
  \bibfield  {author} {\bibinfo {author} {\bibfnamefont {M.~G.~A.}\
  \bibnamefont {Paris}}, \bibinfo {author} {\bibfnamefont {F.}~\bibnamefont
  {Illuminati}}, \bibinfo {author} {\bibfnamefont {A.}~\bibnamefont
  {Serafini}}, \ and\ \bibinfo {author} {\bibfnamefont {S.}~\bibnamefont
  {De~Siena}},\ }\href {\doibase 10.1103/PhysRevA.68.012314} {\bibfield
  {journal} {\bibinfo  {journal} {Physical Review A}\ }\textbf {\bibinfo
  {volume} {68}},\ \bibinfo {pages} {012314} (\bibinfo {year}
  {2003})}\BibitemShut {NoStop}%
\bibitem [{\citenamefont {Bennett}\ \emph {et~al.}(1996)\citenamefont
  {Bennett}, \citenamefont {DiVincenzo}, \citenamefont {Smolin},\ and\
  \citenamefont {Wootters}}]{Bennett1996}%
  \BibitemOpen
  \bibfield  {author} {\bibinfo {author} {\bibfnamefont {C.~H.}\ \bibnamefont
  {Bennett}}, \bibinfo {author} {\bibfnamefont {D.~P.}\ \bibnamefont
  {DiVincenzo}}, \bibinfo {author} {\bibfnamefont {J.~A.}\ \bibnamefont
  {Smolin}}, \ and\ \bibinfo {author} {\bibfnamefont {W.~K.}\ \bibnamefont
  {Wootters}},\ }\href {\doibase 10.1103/physreva.54.3824} {\bibfield
  {journal} {\bibinfo  {journal} {Physical Review A}\ }\textbf {\bibinfo
  {volume} {54}},\ \bibinfo {pages} {3824} (\bibinfo {year}
  {1996})}\BibitemShut {NoStop}%
\bibitem [{\citenamefont {Lloyd}\ and\ \citenamefont
  {Braunstein}(1999)}]{Lloyd1999}%
  \BibitemOpen
  \bibfield  {author} {\bibinfo {author} {\bibfnamefont {S.}~\bibnamefont
  {Lloyd}}\ and\ \bibinfo {author} {\bibfnamefont {S.~L.}\ \bibnamefont
  {Braunstein}},\ }\href {\doibase 10.1103/PhysRevLett.82.1784} {\bibfield
  {journal} {\bibinfo  {journal} {Physical Review Letters}\ }\textbf {\bibinfo
  {volume} {82}},\ \bibinfo {pages} {1784} (\bibinfo {year}
  {1999})}\BibitemShut {NoStop}%
\bibitem [{\citenamefont {Bartlett}\ and\ \citenamefont
  {Sanders}(2002)}]{Bartlett2002}%
  \BibitemOpen
  \bibfield  {author} {\bibinfo {author} {\bibfnamefont {S.~D.}\ \bibnamefont
  {Bartlett}}\ and\ \bibinfo {author} {\bibfnamefont {B.~C.}\ \bibnamefont
  {Sanders}},\ }\href {\doibase 10.1103/PhysRevA.65.042304} {\bibfield
  {journal} {\bibinfo  {journal} {Physical Review A}\ }\textbf {\bibinfo
  {volume} {65}},\ \bibinfo {pages} {042304} (\bibinfo {year}
  {2002})}\BibitemShut {NoStop}%
\bibitem [{\citenamefont {Yanagimoto}\ \emph {et~al.}(2020)\citenamefont
  {Yanagimoto}, \citenamefont {Onodera}, \citenamefont {Ng}, \citenamefont
  {Wright}, \citenamefont {McMahon},\ and\ \citenamefont
  {Mabuchi}}]{Yanagimoto2020}%
  \BibitemOpen
  \bibfield  {author} {\bibinfo {author} {\bibfnamefont {R.}~\bibnamefont
  {Yanagimoto}}, \bibinfo {author} {\bibfnamefont {T.}~\bibnamefont {Onodera}},
  \bibinfo {author} {\bibfnamefont {E.}~\bibnamefont {Ng}}, \bibinfo {author}
  {\bibfnamefont {L.~G.}\ \bibnamefont {Wright}}, \bibinfo {author}
  {\bibfnamefont {P.~L.}\ \bibnamefont {McMahon}}, \ and\ \bibinfo {author}
  {\bibfnamefont {H.}~\bibnamefont {Mabuchi}},\ }\href {\doibase
  10.1103/PhysRevLett.124.240503} {\bibfield  {journal} {\bibinfo  {journal}
  {Physical Review Letters}\ }\textbf {\bibinfo {volume} {124}},\ \bibinfo
  {pages} {240503} (\bibinfo {year} {2020})}\BibitemShut {NoStop}%
\bibitem [{\citenamefont {Petrovnin}\ \emph {et~al.}(2021)\citenamefont
  {Petrovnin}, \citenamefont {Perelshtein}, \citenamefont {Lilja},
  \citenamefont {Korkalainen}, \citenamefont {Vesterinen}, \citenamefont
  {Paraoanu},\ and\ \citenamefont {Hakonen}}]{Petrovnin2021}%
  \BibitemOpen
  \bibfield  {author} {\bibinfo {author} {\bibfnamefont {K.~V.}\ \bibnamefont
  {Petrovnin}}, \bibinfo {author} {\bibfnamefont {M.~R.}\ \bibnamefont
  {Perelshtein}}, \bibinfo {author} {\bibfnamefont {I.}~\bibnamefont {Lilja}},
  \bibinfo {author} {\bibfnamefont {T.}~\bibnamefont {Korkalainen}}, \bibinfo
  {author} {\bibfnamefont {V.}~\bibnamefont {Vesterinen}}, \bibinfo {author}
  {\bibfnamefont {G.~S.}\ \bibnamefont {Paraoanu}}, \ and\ \bibinfo {author}
  {\bibfnamefont {P.~J.}\ \bibnamefont {Hakonen}},\ }\href@noop {} {\bibfield
  {journal} {\bibinfo  {journal} {(to be published)}\ } (\bibinfo {year}
  {2021})}\BibitemShut {NoStop}%
\bibitem [{\citenamefont {Hillmann}\ \emph {et~al.}(2020)\citenamefont
  {Hillmann}, \citenamefont {Quijandr\'{\i}a}, \citenamefont {Johansson},
  \citenamefont {Ferraro}, \citenamefont {Gasparinetti},\ and\ \citenamefont
  {Ferrini}}]{Hillmann2020}%
  \BibitemOpen
  \bibfield  {author} {\bibinfo {author} {\bibfnamefont {T.}~\bibnamefont
  {Hillmann}}, \bibinfo {author} {\bibfnamefont {F.}~\bibnamefont
  {Quijandr\'{\i}a}}, \bibinfo {author} {\bibfnamefont {G.}~\bibnamefont
  {Johansson}}, \bibinfo {author} {\bibfnamefont {A.}~\bibnamefont {Ferraro}},
  \bibinfo {author} {\bibfnamefont {S.}~\bibnamefont {Gasparinetti}}, \ and\
  \bibinfo {author} {\bibfnamefont {G.}~\bibnamefont {Ferrini}},\ }\href
  {\doibase 10.1103/PhysRevLett.125.160501} {\bibfield  {journal} {\bibinfo
  {journal} {Physical Review Letters}\ }\textbf {\bibinfo {volume} {125}},\
  \bibinfo {pages} {160501} (\bibinfo {year} {2020})}\BibitemShut {NoStop}%
\bibitem [{\citenamefont {Samsonov}\ \emph {et~al.}(2020)\citenamefont
  {Samsonov}, \citenamefont {Goncharov}, \citenamefont {Gaidash}, \citenamefont
  {Kozubov}, \citenamefont {Egorov},\ and\ \citenamefont
  {Gleim}}]{Samsonov2020}%
  \BibitemOpen
  \bibfield  {author} {\bibinfo {author} {\bibfnamefont {E.}~\bibnamefont
  {Samsonov}}, \bibinfo {author} {\bibfnamefont {R.}~\bibnamefont {Goncharov}},
  \bibinfo {author} {\bibfnamefont {A.}~\bibnamefont {Gaidash}}, \bibinfo
  {author} {\bibfnamefont {A.}~\bibnamefont {Kozubov}}, \bibinfo {author}
  {\bibfnamefont {V.}~\bibnamefont {Egorov}}, \ and\ \bibinfo {author}
  {\bibfnamefont {A.}~\bibnamefont {Gleim}},\ }\href {\doibase
  10.1038/s41598-020-66948-0} {\bibfield  {journal} {\bibinfo  {journal}
  {Scientific Reports}\ }\textbf {\bibinfo {volume} {10}},\ \bibinfo {pages}
  {10034} (\bibinfo {year} {2020})}\BibitemShut {NoStop}%
\bibitem [{\citenamefont {Guo}\ \emph {et~al.}(2019)\citenamefont {Guo},
  \citenamefont {Breum}, \citenamefont {Borregaard}, \citenamefont {Izumi},
  \citenamefont {Larsen}, \citenamefont {Gehring}, \citenamefont {Christandl},
  \citenamefont {Neergaard-Nielsen},\ and\ \citenamefont {Andersen}}]{Guo2019}%
  \BibitemOpen
  \bibfield  {author} {\bibinfo {author} {\bibfnamefont {X.}~\bibnamefont
  {Guo}}, \bibinfo {author} {\bibfnamefont {C.~R.}\ \bibnamefont {Breum}},
  \bibinfo {author} {\bibfnamefont {J.}~\bibnamefont {Borregaard}}, \bibinfo
  {author} {\bibfnamefont {S.}~\bibnamefont {Izumi}}, \bibinfo {author}
  {\bibfnamefont {M.~V.}\ \bibnamefont {Larsen}}, \bibinfo {author}
  {\bibfnamefont {T.}~\bibnamefont {Gehring}}, \bibinfo {author} {\bibfnamefont
  {M.}~\bibnamefont {Christandl}}, \bibinfo {author} {\bibfnamefont {J.~S.}\
  \bibnamefont {Neergaard-Nielsen}}, \ and\ \bibinfo {author} {\bibfnamefont
  {U.~L.}\ \bibnamefont {Andersen}},\ }\href {\doibase
  10.1038/s41567-019-0743-x} {\bibfield  {journal} {\bibinfo  {journal} {Nature
  Physics}\ }\textbf {\bibinfo {volume} {16}},\ \bibinfo {pages} {281}
  (\bibinfo {year} {2019})}\BibitemShut {NoStop}%
\bibitem [{\citenamefont {Backes}\ \emph {et~al.}(2021)\citenamefont {Backes},
  \citenamefont {Palken}, \citenamefont {Kenany}, \citenamefont {Brubaker},
  \citenamefont {Cahn}, \citenamefont {Droster}, \citenamefont {Hilton},
  \citenamefont {Ghosh}, \citenamefont {Jackson}, \citenamefont {Lamoreaux},
  \citenamefont {Leder}, \citenamefont {Lehnert}, \citenamefont {Lewis},
  \citenamefont {Malnou}, \citenamefont {Maruyama}, \citenamefont {Rapidis},
  \citenamefont {Simanovskaia}, \citenamefont {Singh}, \citenamefont {Speller},
  \citenamefont {Urdinaran}, \citenamefont {Vale}, \citenamefont {van
  Assendelft}, \citenamefont {van Bibber},\ and\ \citenamefont
  {Wang}}]{Backes2021}%
  \BibitemOpen
  \bibfield  {author} {\bibinfo {author} {\bibfnamefont {K.~M.}\ \bibnamefont
  {Backes}}, \bibinfo {author} {\bibfnamefont {D.~A.}\ \bibnamefont {Palken}},
  \bibinfo {author} {\bibfnamefont {S.~A.}\ \bibnamefont {Kenany}}, \bibinfo
  {author} {\bibfnamefont {B.~M.}\ \bibnamefont {Brubaker}}, \bibinfo {author}
  {\bibfnamefont {S.~B.}\ \bibnamefont {Cahn}}, \bibinfo {author}
  {\bibfnamefont {A.}~\bibnamefont {Droster}}, \bibinfo {author} {\bibfnamefont
  {G.~C.}\ \bibnamefont {Hilton}}, \bibinfo {author} {\bibfnamefont
  {S.}~\bibnamefont {Ghosh}}, \bibinfo {author} {\bibfnamefont
  {H.}~\bibnamefont {Jackson}}, \bibinfo {author} {\bibfnamefont {S.~K.}\
  \bibnamefont {Lamoreaux}}, \bibinfo {author} {\bibfnamefont {A.~F.}\
  \bibnamefont {Leder}}, \bibinfo {author} {\bibfnamefont {K.~W.}\ \bibnamefont
  {Lehnert}}, \bibinfo {author} {\bibfnamefont {S.~M.}\ \bibnamefont {Lewis}},
  \bibinfo {author} {\bibfnamefont {M.}~\bibnamefont {Malnou}}, \bibinfo
  {author} {\bibfnamefont {R.~H.}\ \bibnamefont {Maruyama}}, \bibinfo {author}
  {\bibfnamefont {N.~M.}\ \bibnamefont {Rapidis}}, \bibinfo {author}
  {\bibfnamefont {M.}~\bibnamefont {Simanovskaia}}, \bibinfo {author}
  {\bibfnamefont {S.}~\bibnamefont {Singh}}, \bibinfo {author} {\bibfnamefont
  {D.~H.}\ \bibnamefont {Speller}}, \bibinfo {author} {\bibfnamefont
  {I.}~\bibnamefont {Urdinaran}}, \bibinfo {author} {\bibfnamefont {L.~R.}\
  \bibnamefont {Vale}}, \bibinfo {author} {\bibfnamefont {E.~C.}\ \bibnamefont
  {van Assendelft}}, \bibinfo {author} {\bibfnamefont {K.}~\bibnamefont {van
  Bibber}}, \ and\ \bibinfo {author} {\bibfnamefont {H.}~\bibnamefont {Wang}},\
  }\href {\doibase 10.1038/s41586-021-03226-7} {\bibfield  {journal} {\bibinfo
  {journal} {Nature}\ }\textbf {\bibinfo {volume} {590}},\ \bibinfo {pages}
  {238} (\bibinfo {year} {2021})}\BibitemShut {NoStop}%
\bibitem [{\citenamefont {Esposito}\ \emph
  {et~al.}(2021{\natexlab{b}})\citenamefont {Esposito}, \citenamefont
  {Ranadive}, \citenamefont {Planat}, \citenamefont {Leger}, \citenamefont
  {Fraudet}, \citenamefont {Jouanny}, \citenamefont {Buisson}, \citenamefont
  {Guichard}, \citenamefont {Naud}, \citenamefont {Aumentado}, \citenamefont
  {Lecocq},\ and\ \citenamefont {Roch}}]{Esposito2021twpa}%
  \BibitemOpen
  \bibfield  {author} {\bibinfo {author} {\bibfnamefont {M.}~\bibnamefont
  {Esposito}}, \bibinfo {author} {\bibfnamefont {A.}~\bibnamefont {Ranadive}},
  \bibinfo {author} {\bibfnamefont {L.}~\bibnamefont {Planat}}, \bibinfo
  {author} {\bibfnamefont {S.}~\bibnamefont {Leger}}, \bibinfo {author}
  {\bibfnamefont {D.}~\bibnamefont {Fraudet}}, \bibinfo {author} {\bibfnamefont
  {V.}~\bibnamefont {Jouanny}}, \bibinfo {author} {\bibfnamefont
  {O.}~\bibnamefont {Buisson}}, \bibinfo {author} {\bibfnamefont
  {W.}~\bibnamefont {Guichard}}, \bibinfo {author} {\bibfnamefont
  {C.}~\bibnamefont {Naud}}, \bibinfo {author} {\bibfnamefont {J.}~\bibnamefont
  {Aumentado}}, \bibinfo {author} {\bibfnamefont {F.}~\bibnamefont {Lecocq}}, \
  and\ \bibinfo {author} {\bibfnamefont {N.}~\bibnamefont {Roch}},\ }\href@noop
  {} {\enquote {\bibinfo {title} {Observation of two-mode squeezing in a
  traveling wave parametric amplifier},}\ } (\bibinfo {year}
  {2021}{\natexlab{b}}),\ \Eprint {http://arxiv.org/abs/2111.03696}
  {arXiv:2111.03696 [quant-ph]} \BibitemShut {NoStop}%
\bibitem [{\citenamefont {Qiu}\ \emph {et~al.}(2022)\citenamefont {Qiu},
  \citenamefont {Grimsmo}, \citenamefont {Peng}, \citenamefont {Kannan},
  \citenamefont {Lienhard}, \citenamefont {Sung}, \citenamefont {Krantz},
  \citenamefont {Bolkhovsky}, \citenamefont {Calusine}, \citenamefont {Kim},
  \citenamefont {Melville}, \citenamefont {Niedzielski}, \citenamefont {Yoder},
  \citenamefont {Schwartz}, \citenamefont {Orlando}, \citenamefont {Siddiqi},
  \citenamefont {Gustavsson}, \citenamefont {O'Brien},\ and\ \citenamefont
  {Oliver}}]{Qiu2022}%
  \BibitemOpen
  \bibfield  {author} {\bibinfo {author} {\bibfnamefont {J.~Y.}\ \bibnamefont
  {Qiu}}, \bibinfo {author} {\bibfnamefont {A.}~\bibnamefont {Grimsmo}},
  \bibinfo {author} {\bibfnamefont {K.}~\bibnamefont {Peng}}, \bibinfo {author}
  {\bibfnamefont {B.}~\bibnamefont {Kannan}}, \bibinfo {author} {\bibfnamefont
  {B.}~\bibnamefont {Lienhard}}, \bibinfo {author} {\bibfnamefont
  {Y.}~\bibnamefont {Sung}}, \bibinfo {author} {\bibfnamefont {P.}~\bibnamefont
  {Krantz}}, \bibinfo {author} {\bibfnamefont {V.}~\bibnamefont {Bolkhovsky}},
  \bibinfo {author} {\bibfnamefont {G.}~\bibnamefont {Calusine}}, \bibinfo
  {author} {\bibfnamefont {D.}~\bibnamefont {Kim}}, \bibinfo {author}
  {\bibfnamefont {A.}~\bibnamefont {Melville}}, \bibinfo {author}
  {\bibfnamefont {B.~M.}\ \bibnamefont {Niedzielski}}, \bibinfo {author}
  {\bibfnamefont {J.}~\bibnamefont {Yoder}}, \bibinfo {author} {\bibfnamefont
  {M.~E.}\ \bibnamefont {Schwartz}}, \bibinfo {author} {\bibfnamefont {T.~P.}\
  \bibnamefont {Orlando}}, \bibinfo {author} {\bibfnamefont {I.}~\bibnamefont
  {Siddiqi}}, \bibinfo {author} {\bibfnamefont {S.}~\bibnamefont {Gustavsson}},
  \bibinfo {author} {\bibfnamefont {K.~P.}\ \bibnamefont {O'Brien}}, \ and\
  \bibinfo {author} {\bibfnamefont {W.~D.}\ \bibnamefont {Oliver}},\
  }\href@noop {} {\enquote {\bibinfo {title} {Broadband squeezed microwaves and
  amplification with a {J}osephson traveling-wave parametric amplifier},}\ }
  (\bibinfo {year} {2022}),\ \Eprint {http://arxiv.org/abs/2201.11261}
  {arXiv:2201.11261 [quant-ph]} \BibitemShut {NoStop}%
\bibitem [{\citenamefont {Yan}\ \emph {et~al.}(2021)\citenamefont {Yan},
  \citenamefont {Hassel}, \citenamefont {Vesterinen}, \citenamefont {Zhang},
  \citenamefont {Ikonen}, \citenamefont {Grönberg}, \citenamefont {Goetz},\
  and\ \citenamefont {Möttönen}}]{Yan2021}%
  \BibitemOpen
  \bibfield  {author} {\bibinfo {author} {\bibfnamefont {C.}~\bibnamefont
  {Yan}}, \bibinfo {author} {\bibfnamefont {J.}~\bibnamefont {Hassel}},
  \bibinfo {author} {\bibfnamefont {V.}~\bibnamefont {Vesterinen}}, \bibinfo
  {author} {\bibfnamefont {J.}~\bibnamefont {Zhang}}, \bibinfo {author}
  {\bibfnamefont {J.}~\bibnamefont {Ikonen}}, \bibinfo {author} {\bibfnamefont
  {L.}~\bibnamefont {Grönberg}}, \bibinfo {author} {\bibfnamefont
  {J.}~\bibnamefont {Goetz}}, \ and\ \bibinfo {author} {\bibfnamefont
  {M.}~\bibnamefont {Möttönen}},\ }\href@noop {} {\enquote {\bibinfo {title}
  {Low-noise on-chip coherent microwave source},}\ } (\bibinfo {year} {2021}),\
  \Eprint {http://arxiv.org/abs/arXiv 2103.07617} {arXiv:arXiv 2103.07617
  [quant-ph]} \BibitemShut {NoStop}%
\bibitem [{\citenamefont {Frattini}\ \emph {et~al.}(2018)\citenamefont
  {Frattini}, \citenamefont {Sivak}, \citenamefont {Lingenfelter},
  \citenamefont {Shankar},\ and\ \citenamefont {Devoret}}]{Frattini2018}%
  \BibitemOpen
  \bibfield  {author} {\bibinfo {author} {\bibfnamefont {N.~E.}\ \bibnamefont
  {Frattini}}, \bibinfo {author} {\bibfnamefont {V.~V.}\ \bibnamefont {Sivak}},
  \bibinfo {author} {\bibfnamefont {A.}~\bibnamefont {Lingenfelter}}, \bibinfo
  {author} {\bibfnamefont {S.}~\bibnamefont {Shankar}}, \ and\ \bibinfo
  {author} {\bibfnamefont {M.~H.}\ \bibnamefont {Devoret}},\ }\href {\doibase
  10.1103/PhysRevApplied.10.054020} {\bibfield  {journal} {\bibinfo  {journal}
  {Phys. Rev. Applied}\ }\textbf {\bibinfo {volume} {10}},\ \bibinfo {pages}
  {054020} (\bibinfo {year} {2018})}\BibitemShut {NoStop}%
\bibitem [{\citenamefont {Haus}(2000)}]{Haus2000}%
  \BibitemOpen
  \bibfield  {author} {\bibinfo {author} {\bibfnamefont {H.~A.}\ \bibnamefont
  {Haus}},\ }\href@noop {} {\emph {\bibinfo {title} {Electromagnetic Noise and
  Quantum Optical Measurements}}}\ (\bibinfo  {publisher} {Springer Berlin
  Heidelberg},\ \bibinfo {year} {2000})\BibitemShut {NoStop}%
\bibitem [{\citenamefont {Houde}\ \emph {et~al.}(2019)\citenamefont {Houde},
  \citenamefont {Govia},\ and\ \citenamefont {Clerk}}]{Houde2019}%
  \BibitemOpen
  \bibfield  {author} {\bibinfo {author} {\bibfnamefont {M.}~\bibnamefont
  {Houde}}, \bibinfo {author} {\bibfnamefont {L.}~\bibnamefont {Govia}}, \ and\
  \bibinfo {author} {\bibfnamefont {A.}~\bibnamefont {Clerk}},\ }\href
  {\doibase 10.1103/PhysRevApplied.12.034054} {\bibfield  {journal} {\bibinfo
  {journal} {Phys. Rev. Applied}\ }\textbf {\bibinfo {volume} {12}},\ \bibinfo
  {pages} {034054} (\bibinfo {year} {2019})}\BibitemShut {NoStop}%
\bibitem [{\citenamefont {Flurin}\ \emph {et~al.}(2015)\citenamefont {Flurin},
  \citenamefont {Roch}, \citenamefont {Pillet}, \citenamefont {Mallet},\ and\
  \citenamefont {Huard}}]{Flurin2015}%
  \BibitemOpen
  \bibfield  {author} {\bibinfo {author} {\bibfnamefont {E.}~\bibnamefont
  {Flurin}}, \bibinfo {author} {\bibfnamefont {N.}~\bibnamefont {Roch}},
  \bibinfo {author} {\bibfnamefont {J.~D.}\ \bibnamefont {Pillet}}, \bibinfo
  {author} {\bibfnamefont {F.}~\bibnamefont {Mallet}}, \ and\ \bibinfo {author}
  {\bibfnamefont {B.}~\bibnamefont {Huard}},\ }\href {\doibase
  10.1103/PhysRevLett.114.090503} {\bibfield  {journal} {\bibinfo  {journal}
  {Phys. Rev. Lett.}\ }\textbf {\bibinfo {volume} {114}},\ \bibinfo {pages}
  {090503} (\bibinfo {year} {2015})}\BibitemShut {NoStop}%
\bibitem [{\citenamefont {Laurat}\ \emph {et~al.}(2005)\citenamefont {Laurat},
  \citenamefont {Coudreau}, \citenamefont {Keller}, \citenamefont {Treps},\
  and\ \citenamefont {Fabre}}]{Laurat2005}%
  \BibitemOpen
  \bibfield  {author} {\bibinfo {author} {\bibfnamefont {J.}~\bibnamefont
  {Laurat}}, \bibinfo {author} {\bibfnamefont {T.}~\bibnamefont {Coudreau}},
  \bibinfo {author} {\bibfnamefont {G.}~\bibnamefont {Keller}}, \bibinfo
  {author} {\bibfnamefont {N.}~\bibnamefont {Treps}}, \ and\ \bibinfo {author}
  {\bibfnamefont {C.}~\bibnamefont {Fabre}},\ }\href {\doibase
  10.1103/PhysRevA.71.022313} {\bibfield  {journal} {\bibinfo  {journal}
  {Physical Review A}\ }\textbf {\bibinfo {volume} {71}},\ \bibinfo {pages}
  {022313} (\bibinfo {year} {2005})}\BibitemShut {NoStop}%
\bibitem [{\citenamefont {Flurin}\ \emph {et~al.}(2012)\citenamefont {Flurin},
  \citenamefont {Roch}, \citenamefont {Mallet}, \citenamefont {Devoret},\ and\
  \citenamefont {Huard}}]{Flurin2012}%
  \BibitemOpen
  \bibfield  {author} {\bibinfo {author} {\bibfnamefont {E.}~\bibnamefont
  {Flurin}}, \bibinfo {author} {\bibfnamefont {N.}~\bibnamefont {Roch}},
  \bibinfo {author} {\bibfnamefont {F.}~\bibnamefont {Mallet}}, \bibinfo
  {author} {\bibfnamefont {M.~H.}\ \bibnamefont {Devoret}}, \ and\ \bibinfo
  {author} {\bibfnamefont {B.}~\bibnamefont {Huard}},\ }\href {\doibase
  10.1103/PhysRevLett.109.183901} {\bibfield  {journal} {\bibinfo  {journal}
  {Physical Review Letters}\ }\textbf {\bibinfo {volume} {109}},\ \bibinfo
  {pages} {183901} (\bibinfo {year} {2012})}\BibitemShut {NoStop}%
\end{thebibliography}%

\end{document}